\documentclass[journal]{IEEEtran}

\usepackage[thmmarks]{ntheorem}

\usepackage{cite}
\usepackage{enumitem}
\usepackage{ragged2e}
\usepackage{algorithm}
\usepackage{algorithmic}
\ifCLASSINFOpdf
\usepackage[pdftex]{graphicx}
 \graphicspath{{../pdf/}{../jpeg/}}
 \DeclareGraphicsExtensions{.pdf,.jpeg,.png}
\else
 \usepackage[dvips]{graphicx}
  \graphicspath{{../eps/}}
\DeclareGraphicsExtensions{.eps}
\fi
\usepackage{amsmath}
\usepackage{amssymb}
\usepackage{array}
\ifCLASSOPTIONcompsoc
  \usepackage[caption=false,font=normalsize,labelfont=sf,textfont=sf]{subfig}
\else
 \usepackage[caption=false,font=footnotesize]{subfig}
\fi

\usepackage{subfig}
\usepackage{stfloats}
\usepackage{footmisc}
\usepackage{tabu}
\usepackage{multirow}
\usepackage{xcolor}

\ifCLASSOPTIONcaptionsoff
\usepackage[nomarkers]{endfloat}
\let\MYoriglatexcaption\caption
 \renewcommand{\caption}[2][\relax]{\MYoriglatexcaption[#2]{#2}}
\fi
\usepackage{url}
\newcommand{\tabitem}{\makebox[1em][r]{\textbullet~}}
\makeatletter
\def\blfootnote{\xdef\@thefnmark{}\@footnotetext}
\makeatother

\begin{document}
\title{A Hierarchical Performance Equation Library for Basic Op-Amp Design}

\author{Inga Abel, Maximilian Neuner,
        and~Helmut~Graeb~
\\ Technical University of Munich, Chair of Electronic Design Automation, $\{$first name$\}$.$\{$last name$\}$@tum.de}

\maketitle

\begin{abstract}
The paper presents a new approach to automate the set-up of the design equations of the manual analog design process.
Its main contribution is a comprehensive hierarchical performance equation library (HPEL) for op-amps. 
The HPEL makes the set-up of design equations independent of the topology.
Based on the library and the functional block recognition method in \cite{ABNG20}, analytical performance models for various op-amp topologies are automatically instantiated.  
The method is currently designed for basic op-amps. 
In this paper, we use the method to size different op-amp topologies. Experimental results featuring four circuits are presented.
The HPEL has also been integrated into a structural synthesis method featuring several thousand op-amp topologies \cite{ABNG20d}.
\end{abstract}

\IEEEpeerreviewmaketitle

\section{Introduction}
{\blfootnote{\copyright 2021 IEEE. Personal use is permitted, but republication/redistribution requires IEEE permission.
		See https://www.ieee.org/publications/rights/index.html for more information.
		This article has been accepted for publication in IEEE Transactions on Computer-Aided Design of Integrated Circuits and Systems. This is the author's version which has not been fully edit
		content may change prior to final publication. Citation information: DOI 10.1109/TCAD.2021.3101691}
}

Behavioral equations are a major means in the design process of analog circuits to analyze the DC-, AC-, and transient behavior of the circuit without requiring time-consuming circuit simulation.
They are used for instance for structural synthesis and in the sizing process.

Structural synthesis aims at finding a suitable netlist of transistors (topology) for a given set of specifications. Behavioral equations are used to guide the topology selection and development process, e.g.~\cite{IntegerBasedTopologieSelecting, NovelCircuitTopologySynthesisMethodUsingCircuitFeatureMiningAndSymbolicComparison,AGenericTopologySelectionMethodForAnalogCircuits,HARC89,VariationAwareStructuralSynthesisOfAnalogCircuitsViaHierarchicalBuildingBlocksAndStructuralHomotopy,AnAutomatedTopologySynthesisFrameworkForAnalogIntegratedCircuits}.

Sizing is the process of finding the device sizes in an analog circuit, e.g., the widths and lengths of CMOS transistors, such that the performance specifications, e.g., for gain, power consumption, slew rate, are fulfilled.
Many computer-aided approaches for sizing are equation-based, 
e.g.,~\cite{DEGR87,ELPE89,KOSG90a,MACA91,GPCAD,LDGS97a,PDLL01,SHTA00,VEGI01,LPRT13,ICCD2019,ABNG20b}.
Simulation-based sizing approaches, e.g.~\cite{DelightSPICE,ANGR91a,ANGW94,OCRC96,PKRC00,LIFG11,MARS,GYSH14,BADP15,PRSF19,CPML19,FEATS},
use numerical SPICE-like simulation. 
They are considered an alternative to or an afterburner of equation-based sizing. They deal with any type circuit, but are computationally more expensive. The optimizer and suitable constraints must be set up manually, a performance evaluation by numerical simulation is more expensive than by analytical equations and the numerical optimization process during sizing is more difficult to understand from the physical point of view.

Analog designers therefore often prefer equation-based sizing approaches.
Notwithstanding the increasing significance of simulation-based analog circuit design in deep 
submicron process technologies, an initial sizing based on analytical equations is the gold
standard in analog design. It makes numerical performance evaluation within the sizing process unnecessary saving computational cost and being closer to the designer wish for physical insights.
However to automatize the initial sizing process, the major obstacle is the automatic set-up of the design equations. This is where this paper presents a new approach.

\begin{table} [] \scriptsize\centering
	\caption{Comparison of the state-of-the-art equation-based sizing tools with the here presented method}\label{tab:MethodComparsion}
	\begin{tabu}{|>{\raggedright\arraybackslash}m{1.9cm}|>{\centering\arraybackslash}m{1.4cm}|>{\centering\arraybackslash}m{2cm}|>{\centering\arraybackslash}m{1.3cm}|}
		\hline
		Tool/author & Supported circuits  &  Equation set-up  & Set-up time new circuit \\\hline
		IDAC~\cite{DEGR87} & amplifiers  & topology database& {months} \\
		BLADES~\cite{ELPE89} & op-amps, subblocks & subblock-based&{\bf }long  \\
		OASYS~\cite{HARC89} & op-amps & topology database & month \\
		OPASYN~\cite{KOSG90a} & op-amps & topology database& few weeks        \\
		Maulik~ \cite{MACA91} & op-amps & topology database & long \\
		GPCAD~\cite{GPCAD} & op-amps & topology database & { \bf} long\\\hline
		Leyn~ \cite{LDGS97a} & amplifiers & symbolic analysis & hours \\
		AMGIE~ \cite{PDLL01} & op-amps & topology database, symbolic analysis & 8 h \\
		Shi~ \cite{SHTA00} & op-amps& symbolic analysis & seconds \\
		Verhagen~ \cite{VEGI01} & op-amps& symbolic analysis & seconds  \\
		Liu~ \cite{LPRT13} & op-amps &  symbolic analysis & seconds   \\\hline
		COPRICSI~ \cite{ABNG20b} & op-amps & subblock-based & few days  \\
		This paper: HPEL & op-amps & subblock-based &  seconds(*)  \\\hline
     \multicolumn{4}{|l|}{(*): new subblock 0.5-2 days} \\\hline
	\end{tabu}	
\end{table}

Equation-based synthesis or sizing approaches (Table~\ref{tab:MethodComparsion})  have presented fixed design plans for supported process technologies~\cite{DEGR87,ELPE89,HARC89,KOSG90a, MACA91, GPCAD}, or apply symbolic analysis to create transfer functions automatically~\cite{LDGS97a, PDLL01,SHTA00,VEGI01,LPRT13}.
The methods support mainly op-amps \cite{ELPE89, KOSG90a, MACA91, GPCAD,PDLL01,SHTA00,VEGI01, LPRT13, ICCD2019, ABNG20b} and  other types of amplifiers \cite{DEGR87,LDGS97a}. 

Early equation-based methods \cite{DEGR87,KOSG90a,MACA91,GPCAD, HARC89} stored the equations topology-dependent. Topology libraries were developed containing a fixed equation set for every supported topology.  To overcome the topology dependence, \cite{ELPE89} splits  up part of the equation-based description into subcircuits descriptions. However, only basic equations, as symmetry constraints and DC-performance constraints, are considered for the subcircuits. AC- and transient performance constraints are still restricted to a specific topology. 
Adding new topologies to these methods takes quite long as for every topologies a new equation-based description must be developed.

To reduce the set-up time of the equation-based description, symbolic analysis method were developed \cite{LDGS97a, PDLL01, SHTA00, VEGI01, LPRT13}. They automatically create the transfer function of a given topology. The transfer function, however,  only represents the AC-behavior of the circuit. To represent the transient and DC-behavior of the circuit, other methods are still needed. Designer knowledge is for instance used in \cite{PDLL01} to include performance features for the transient behavior and symmetry constraints in the equation-based model. 

To overcome the topology dependence, \cite{ABNG20b} presented a building block analysis to set up part of the equation-based description automatically. However, a method which allows an automatic set-up of the whole equation-based description of the circuit using the same equations as in the manual design process has not been published yet.

Such a method, called hierarchical performance equation library (HPEL), is presented in this paper. HPEL allows the automatic set-up of an equation-based description of a topology emulating the manual design process. AC-, DC- and transient behavior of a circuit are modeled in the same way as in the manual analog design process instantiating well-known model equations \cite{LakerSansen, Allen, AnalogIntegratedCircuitDesign, MOSCapacitances} automatically. The method supports many different op-amp topologies including several thousands of topology variants \cite{ABNG20d}. 

The main contributions of this paper are:
\begin{itemize}
	\item A functional block-based hierarchical generic equation library (Sec.~\ref{sec:HPEL} - Sec.~\ref{sec:opAmpPerformanceConstraints}), storing an equation set for every functional block in \cite{ABNG20} which describes its behavior within the circuit. Compared to the state of the art, the equations are presented comprehensively, not in excerpts. The equations include hierarchically built performance equations and symmetry constraints automatically generated based on the hierarchy of a given circuit. The equation library provides a computer-oriented and hierarchical systematic of the behavior along the hierarchy  the functional block composition of a circuit.
	\item Algorithms to automatically instantiated a behavior circuit model for a given netlist (Sec. \ref{sec:AutomaticModelCreation}). The generic equations in the library are automatically specified for a given topology.
	A nodal analysis model from Kirchhoff voltage and current laws analogous to circuit simulation  leads to a topology-independent set-up of the comprehensive behavioral model for a given circuit netlist. Setting up the problem for a new circuit that is covered by the available equation library takes a few seconds. The circuit model can be fed into a constrained optimization solver for a fast sizing that mimics the sizing approach preferred by analog designers in practice.
\end{itemize}
The hierarchical character of the performance equation library along the functional block composition of a circuit represents a new level of generalization in equation-based design.
The functional blocks and their behavioral models are general modules in analog design.
They are used in advanced op-amps and other circuit classes. 
If such advanced design concepts for op-amps and other circuit classes are to be investigated, the performance library is not set up from scratch, but re-used and extended for new functional blocks only. Depending on the amount of additional new functionality, this is estimated to take between half a day and two days.
Hence, the (manual) inclusion of new topologies into the HPEL library is fast compared to the state of the art as equation sets of functional blocks from lower levels are re-used. 
In topology-based approaches as, e.g., \cite{PDLL01}, the set-up time refers to one specific topology. Adding new functional blocks to the HPEL, however, means the inclusion of whole sets of topologies.

An application of HPEL is a sizing process described in Sec.~\ref{sec:InitialSizingTool}. Circuits are sized in one minute without requiring much manual set-up. The HPEL sets up all constraints required for sizing fully automatically. This is different to numerical sizing approaches, which require a manual set-up of simulation, waveform postprocessing, parameters and constraints which takes around half a day of time.

Experimental results (Sec.~\ref{sec:ExperimentalResults}) present the circuit behavior models established through HPEL for four different circuits. Additionally, sizing results are presented obtained by the performance models.

\section{Functional Blocks in Op-Amps}\label{sec:functionalBlocks}
Every op-amp consists of a set of transistor blocks which can be characterized by their function and are called functional blocks in the following. These functional blocks can be hierarchically structured (Fig. \ref{fig:FunctionalBlockLibrary}). 
With every hierarchy level, the structural composition of the functional block becomes less definable however its overall function in the circuit becomes more clear.
The functional block types on every hierarchy level are sketched in the following. 
A complete description with structural examples is given in \cite{ABNG20}.

\begin{figure}
\centering \scriptsize
	\setlength{\tabcolsep}{0.1cm}
	\begin{tabular}{l>{\arraybackslash}m{2cm}>{\arraybackslash}m{5.4cm}}
		HL 1: & Devices & Normal transistors ($nt$), diode transistors ($dt$), capacitors ($cap$) \\ \hline
		HL 2: & Structures & Voltage biases ($vb$), current biases ($cb$),  current mirror ($cm$), differential pairs ($dp$), analog inverter ($inv$) \\\hline
		HL 3: & Amplification stage subblocks & Transconductor ($tc$), load ($l$), stage bias ($b_s$) \\\hline
		HL 4: & Op-amp subblocks & Amplification stage ($a$), circuit bias ($b_O$), compensation ($c_C$) and load capacitor ($c_L$)\\\hline
		HL 5: & Op-amps & Miller op-amp, Folded-cascode op-amp \\
	\end{tabular}
	\caption{Functional blocks in op-amps}\label{fig:FunctionalBlockLibrary}	
\end{figure}

Hierarchy level 1 consists of devices, e.g., {\em capacitors}  (Fig.~\ref{fig:TelescopicOpAmp} $cap$), and transistors. Two types of transistors are distinguished by their self-connections.
{\em Normal transistors} ($nt$) do not have any self-connection, e.g., Fig. \ref{fig:TelescopicOpAmp}, $nt_1$. {\em Diode transistors} ($dt$) have a gate-drain connection, e.g., Fig.~\ref{fig:TelescopicOpAmp} $dt_1$.

{ Hierarchy level 2} consists of transistor structures: {\em voltage bias} $vb_k$ (Fig.~\ref{fig:TelescopicOpAmp} $vb_1,vb_3$), {\em current bias} $cb_k$ (Fig.~\ref{fig:TelescopicOpAmp} $cb_1, cb_6$), {\em analog inverter} $inv_k$ (Fig.~\ref{fig:TelescopicOpAmp} $inv_1$), {\em differential pair}  (Fig.~\ref{fig:symmetricalOpAmp} $N_1,N_2$).
A voltage bias and a current bias may form a {\em current mirror} ($cm_k$), e.g., Fig. \ref{fig:TelescopicOpAmp} $cm_5$. However, cases exist where no current mirrors are formed \cite{ABNG20}, e.g., Fig. \ref{fig:TelescopicOpAmp}, $vb_2, cb_4$.
Types of differential pairs are: {\em simple} ($dp_k$), {\em cascode} ($cdp_k$), e.g., Fig.~\ref{fig:TelescopicOpAmp} $P_1 - P_4$, or {\em folded-cascode} ($fcdp$), e.g. Fig.~\ref{fig:foldedCascodeOpAmp} $N_1, N_2, P_1, P_2$. A cascode or folded-cascode differential pair consists of a simple differential pair connected to a {\em gate-connected couple} ($gcc$, e.g., Fig.~\ref{fig:TelescopicOpAmp} $P_3,P_4$; Fig. \ref{fig:foldedCascodeOpAmp} $P_1, P_2$).

Hierarchy level 3 consists of the amplification stage subblocks, which are the {\em transconductor} $tc$, the {\em load} $l$ and the {\em stage bias} $b_s$. 
For the transconductor, two main  types exist: non-inverting $tc_{ninv}$ (Fig.~\ref{fig:symmetricalOpAmp} $tc_1$) and inverting $tc_{inv}$ (Fig.~\ref{fig:symmetricalOpAmp} $tc_{2,1}$).
The non-inverting transconductor is  further divided into three types: simple  $tc_s$ (Fig. \ref{fig:symmetricalOpAmp} $tc_1$), complementary $tc_c$, (Fig.~\ref{fig:railToRailAmplifier} $tc_1$)  and common-mode feedback (CMFB) $tc_{CMFB}$ (Fig. \ref{fig:foldedCascodeOpAmp} $tc_{CMFB}$).  The load consists of one or two load parts ($l_p$) (Fig. \ref{fig:foldedCascodeOpAmp}).
The  stage bias is either of type current bias (Fig~\ref{fig:symmetricalOpAmp}, $b_{s,1}$) or of type voltage bias (Fig \ref{fig:symmetricalOpAmp}, $b_{s,2,2}$).

Hierarchy level 4 consists of the op-amp subblocks which are the {\em amplification stages} $a$, the {\em circuit bias} $b_O$ (Fig. \ref{fig:railToRailAmplifier}), the {\em compensation capacitor} ($c_C$) (Fig. \ref{fig:symmetricalOpAmp}), and  {\em load capacitor} ($c_L$).
Two types of amplification stages exist: non-inverting $a_{ninv}$ (Fig. \ref{fig:symmetricalOpAmp} $a_1$), and inverting $a_{inv}$ (Fig. \ref{fig:symmetricalOpAmp} $a_{2,2}$).  The non-inverting amplification stage can be further divided into simple  $a_s$ (Fig. \ref{fig:symmetricalOpAmp} $a_1$) and complementary first stage $a_c$ (Fig. \ref{fig:railToRailAmplifier} $a_1$), and common mode feedback stage $a_{CMFB}$ (Fig. \ref{fig:foldedCascodeOpAmp} $a_{CMFB}$).

Hierarchy level 5 consists of the op-amp itself.
It is fully-differential or has a single output.

The functional blocks in an op-amp are identified as detailed in \cite{ABNG20}. It uses a formalized structural definition of every functional block for an automatic recognition. Starting on the lowest hierarchy level, the functional blocks are hierarchically identified by analyzing the pin connections in the circuit netlist based on the structural definitions of the functional blocks. The result are the specific functional blocks of a given netlist according to the hierarchy levels in Fig.~\ref{fig:FunctionalBlockLibrary}. Examples are the decompositions in Fig.~\ref{fig:differentOpAmpToplogies}. It is worth noting that the functional attribution of a group of transistors depends on its context, i.e., its connection to other circuit parts. The respective performance equations of each functional block type and their automatic set-up are presented in the following.

\section{Overview of the Hierarchical Performance Equation Library}\label{sec:HPEL}
The hierarchical performance equation library uses the functional block description of op-amps to store the equation describing the op-amp behavior topology independent. It distinguishes between two main groups of equations: the basic model and the op-amp performance model.

\begin{table*}\centering\scriptsize
	\setlength{\tabcolsep}{0.1cm}	
	\caption{Hierarchical performance equation library}\label{tab:opAmpModelLibrary}	
	\begin{tabular}{|>{\RaggedRight\arraybackslash}p{2.1cm}|>{\RaggedRight\arraybackslash}p{3.3cm}|>{\RaggedRight\arraybackslash}p{4cm}|>{\RaggedRight\arraybackslash}p{3.7cm}|>{\RaggedRight\arraybackslash}p{3.5cm}|}
		\hline
		&Symmetry constraints & Functional block behavioral constraints & Intermediate performance equations & Op-amp performance equations\\\hline
		HL 1:  Devices &  & &  	\tabitem Saturation drain-source voltage \linebreak \tabitem Net capacitance &  	\tabitem Area \linebreak \tabitem Quiescent power 	\\ \hline
		HL 2:  Structures & \tabitem Voltage and current bias  & \tabitem Current mirror behavior  & & \\\hline
		HL 3:  Amplification stage subblocks & \tabitem Load \linebreak \tabitem Non-inverting transconductor  & \tabitem Complementary transconductor and stage bias  & \tabitem Transconductance \linebreak \tabitem Output conductance  &  \\\hline
		HL 4:\linebreak Op-amp subblocks & \tabitem Inverting stages & \tabitem Output voltage offset & \tabitem Stage output resistance \linebreak \tabitem Stage open-loop gain \linebreak \tabitem Stage non-dominate poles \linebreak \tabitem Stage zeros & \tabitem Common-mode input voltage \linebreak \tabitem Output voltage \linebreak \tabitem Common-mode rejection ratio \linebreak \tabitem Unity-gain bandwidth \\\hline
		HL 5:  Op-amps & & & \tabitem Dominant pole \linebreak \tabitem Positive zero & \tabitem Open-loop gain \linebreak \tabitem Slew Rate \linebreak \tabitem Phase margin \\\hline
	\end{tabular}
	
\end{table*} 

The {\em basic model} describes the current and voltage flow in the circuit. It contains information gained based on the circuit netlist and an analysis of its devices. It comprises the variables of the circuit, Kirchoffs Current and Voltage Law and models for the devices. The variables can be reduced by using information from higher functional block levels, however the major set-up is based on the first hierarchy level.

The {\em op-amp performance model} describes the AC-, DC-, and transient behavior of the op-amp. It  contains information gained from the hierarchical composition of functional blocks. It comprises symmetry constraints, functional block constraints, intermediate performance equations and op-amp performance equations. 
An overview of the op-amp performance model part of the hierarchical equation library is given in Table~\ref{tab:opAmpModelLibrary}. For each hierarchy level, the most important equations or constraints are given. The ordering from left to right represents an abstraction from constraints to performance and it corresponds to the functional abstraction from top to bottom through the hierarchy levels. 
Equations based on functional blocks of low hierarchy levels highly depend on the transistor structure, e.g., the  equations to describe the output conductance of a functional block. To set up the equations of the open-loop gain, the transistor structure of op-amp is ignored. This hierarchical structuring allows us to  generalize the op-amp equations such that we obtain an automatic set-up of the design equations independent of the topology (Sec. \ref{sec:AutomaticModelCreation}).

In the following, variables and equations of both model types are described in detail.

\section{Variables}\label{sec:variables}

The variables of the equation-based topology description can be divided into two groups: device specific variables and  op-amp performance variables.

\emph{Device specific variables:}
Depending on the device type, a set of variables is automatically derived. For a transistor $t_k$, this set is:
\begin{equation}\label{eq:transistorVariable}
	{\bf t}_k^T = \{w_k, l_k, gm_k, gd_k, i_{DS, k}, v_{GS,k}, v_{DS,k}\}
\end{equation} 
$w_k, l_k$ are the width and length of the transistor, $gm_k, gd_k$ its transconductance and output conductance,  $i_{DS,k}$ its drain-source current and $v_{GS,k}, v_{DS,k}$ its gate-source and drain-source voltage.

\emph{Op-amp performance variables:}
The set of characteristic performance features whose equations are automatically set up based on the HPEL is:
\begin{equation}\label{eq:performanceVariable}
\begin{split}
{\bf z}^T =  \{& z_D, z_{QP}, z_{v_{cm,min/max}}, z_{v_{out,min/max}}, \\
&z_{f_{GBW}}, z_{SR}, z_{A_{D0}}, z_{CMRR}, z_{PM}\}
\end{split}
\end{equation}
$z_D$ describes the gate area of the circuit, $z_{QP}$ its quiescent power, $z_{v_{cm,min/max}}$ is the minimal, respective maximal common-mode input voltage, $z_{v_{out,min/max}}$ the minimal/maximal output voltage of the op-amp,  $z_{A_{D0}}$ is its open-loop gain. $z_{f_{GBW}}$ is the unity-gain bandwidth, $z_{SR}$ the slew rate,  $z_{CMRR}$  the common-mode rejection ratio, $z_{PM}$ is the phase margin.
The performance features describe the characteristic op-amp behavior.
Additional to them, intermediate performance variables, e.g., output resistance $R_{out, a_j}$ of a stage $a_j$ or  poles and zeros of an op-amp $\mathbf{z}_{P}$ exist.

\section{Kirchoffs Current and Voltage Law}\label{sec:KCL}

By automatically analyzing the graph description of the netlist, Kirchoffs Current Law (KCL) is set up for every node $l \in \mathcal{N}$ in the circuit:
\begin{equation}
 \forall_{l \in \mathcal{N}} \sum_{k} i_{DS_k}=0 ,  
\end{equation}

Kirchoffs Voltage Law is expressed efficiently by the voltage potentials of the circuit nodes as in circuit simulation.
The voltage variables are therefore all $n_N$ node voltages $\mathbf{v}_{N}$ :
\begin{equation}\label{eq:KVL}
\begin{split}
&\mathbf{v}_{N}^T = [v_{N,1}, v_{N,2}, ... , v_{N,n_N} ], v_{N,k} \in \mathbb{R}, k = 1,2,..., n_N 
\\
&[ \mathbf{v}_{GS}^T  \mathbf{v}_{DS}^T  ]^T    = \mathbf{A} \cdot \mathbf{v}_{N}\text{, with $\mathbf{A}$ as nodal incidence matrix.}   
\end{split}
\end{equation}

\section{Transistor Behavior Model}\label{sec:transistorBehaviorModel}

For every device in the circuit, a model is needed which describes its behavior depending on its variables. Therefore, for every device type on level 1, a behavioral model is stored in the equation library.
For transistors, this is the Shichman-Hodges model \cite{Shichman1968}. It is the simplest transistor model and defines three operating regions for a transistor, off, linear, and saturation. Analytical equations exist for all operating regions. 
The drain-source current of an nmos transistor $t_k$  in saturation is for instance described by:
\begin{equation}
i_{DS,k} = \frac{\mu_k C_{ox,k}}{2} \frac{W_k}{L_k}(v_{GS,k}- v_{th,k})^2 (1+ \lambda_k v_{DS,k} )
\end{equation}
Process parameters, e.g., threshold voltage $v_{th}$, are specified by the underlying process technology. 
Equations for other operation regions are implemented analogously.

Constraints for the transistor voltages  guarantee that the transistor operates in the specified region.
For saturation, these are:
\begin{equation}\label{eq:voltageConstraints}
\begin{split}
v_{GS_,k} - v_{th,k} \geq 0\\
v_{GS_,k} - v_{th,k} < v_{DS,k}
\end{split}
\end{equation} 

The  transconductance and output conductances $gm_{k}, gd_{k}$ of a transistor are calculated by the differentiation of the drain-source current with respect to the transistor voltages.
For the saturation region, following equations are obtained:
\begin{equation}
gm_{k} = \frac{\partial{i_{DS,k}}}{\partial{v_{GS,k}}} = \sqrt{2\mu_k C_{ox,k} \frac{W_k}{L_k} i_{DS,k}}
\end{equation}
\begin{equation}
gd_{k} = \frac{\partial{i_{DS,k}}}{\partial{v_{DS,k}}} = \lambda_k \cdot i_{DS,k}
\end{equation}

The saturation region can be further divided into weak, moderate and strong inversion. An overview how the three inversions region are integrated into equation-based modeling is given in \cite{ANALOG2020}.

\begin{figure*} [tb] \centering
	\subfloat[Telescopic op-amp {[colored background: functional blocks of HL 1 - 2]}]{
		\label{fig:TelescopicOpAmp}
		\includegraphics[width= 0.35\linewidth]{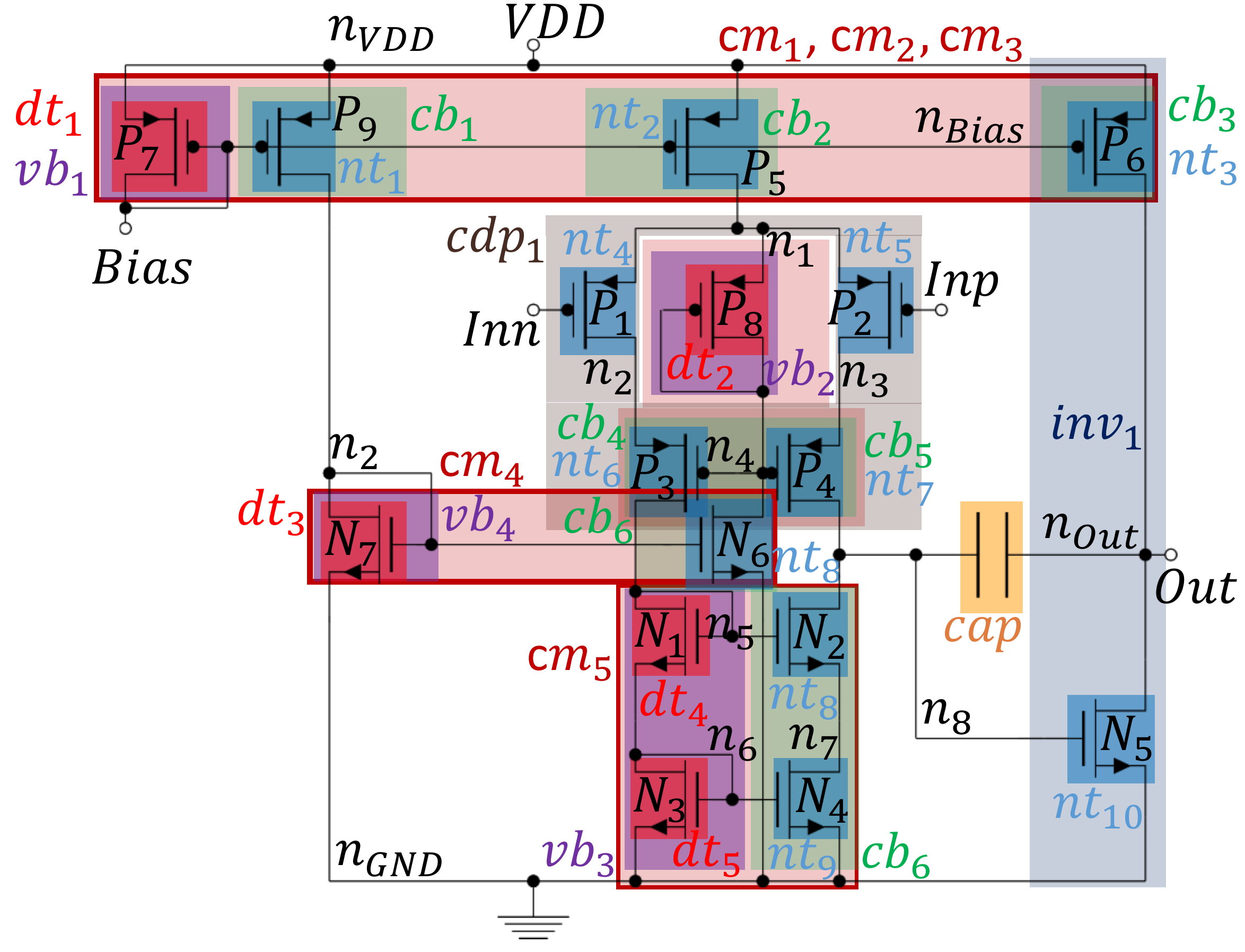}
	}%
	\qquad
	\subfloat[Folded-cascode op-amp with CMFB {[colored background: functional blocks of HL 3 - 4]}]{
		\label{fig:foldedCascodeOpAmp}
		\includegraphics[width=0.5\linewidth]{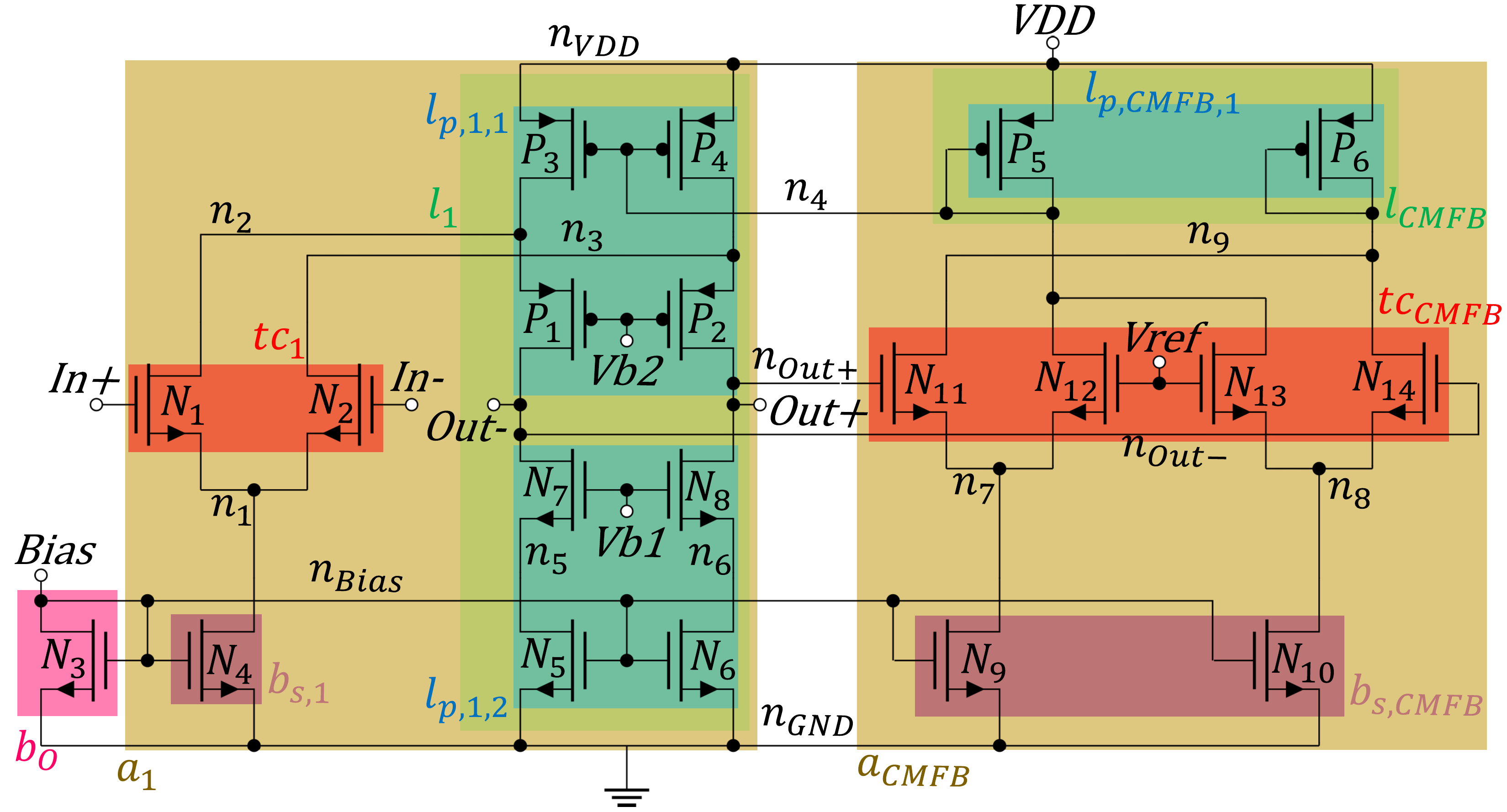}
	}%
	\qquad
	\subfloat[Symmetrical op-amp with high PSRR \cite{LakerSansen} {[colored background: functional blocks of HL 3 - 4]}]{
	\label{fig:symmetricalOpAmp}
	\includegraphics[width=0.35\linewidth]{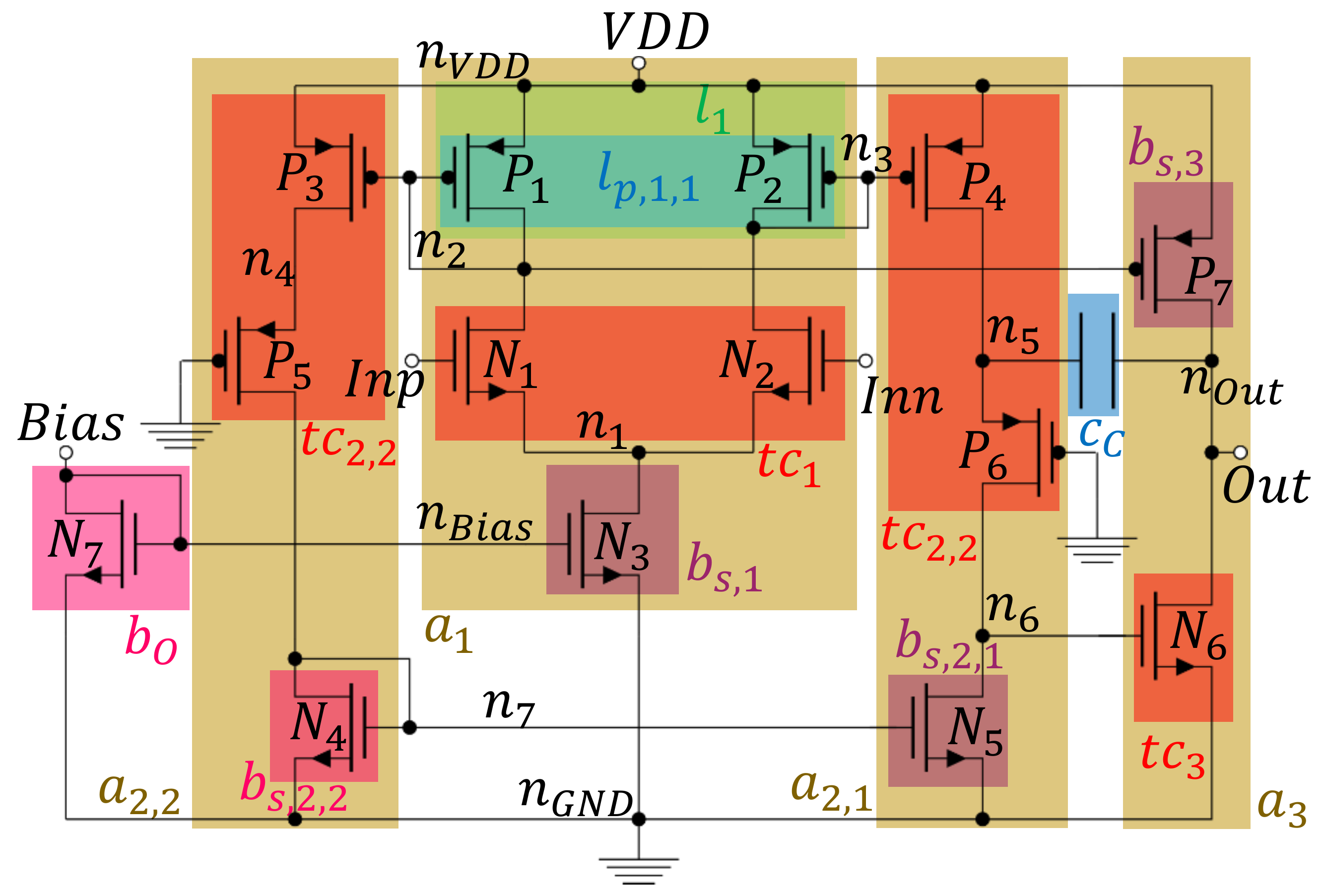}
	}%
	\qquad
	\subfloat[Complementary op-amp {[colored background: functional blocks of HL 3 - 4]}]{
	\label{fig:railToRailAmplifier}
	\includegraphics[width=0.5\linewidth]{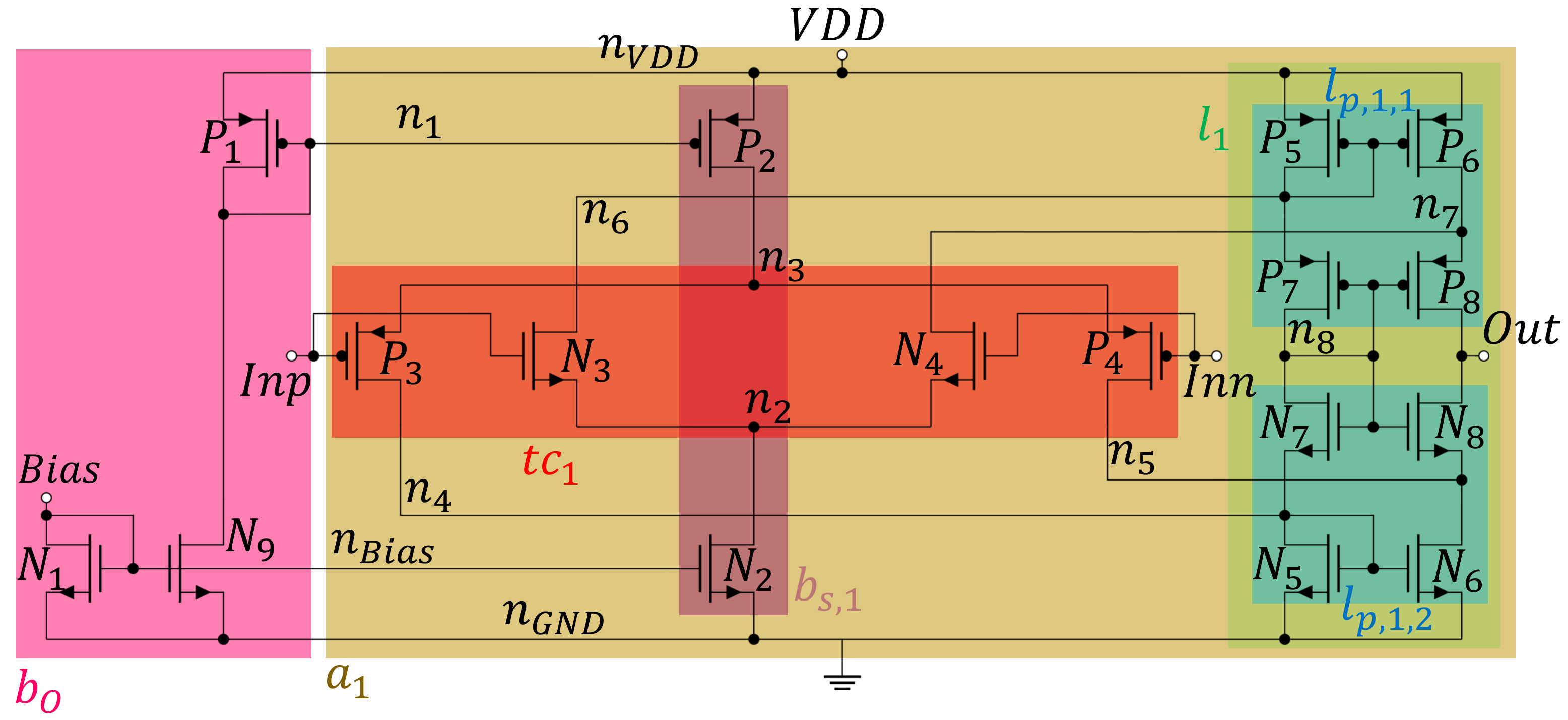}
	}%
	\qquad

	\caption{Different op-amp topologies}
	\label{fig:differentOpAmpToplogies}
\end{figure*}

\section{Symmetry Constraints}\label{sec:symmetryConstraints}
Symmetry constraints are crucial in analog circuits to
minimize mismatch, e.g., due to channel length modulation or local manufacturing variations. Symmetry constraints are derived for structures (HL 2), subblocks of amplification stages (HL~3) and op-amp subblocks (HL 4).
They reduce the number of variables of the performance model.

\subsection{Hierarchy Level 2: Structures}
For transistors in a voltage or current bias, we  define that two transistors $t_i,t_j$ connected at their gates $t_i.g,t_j.g$ must have equal lengths $l_{t_i}, l_{t_j}$
\begin{equation}\label{eq:symVbCb}
\forall_{t_i,t_j \in ({T}_{vb, \Phi} \cup T_{cb, \Phi})} t_i.g \leftrightarrow t_j.g \Rightarrow l_{t_i} = l_{t_j}
\end{equation}
$T_{vb, \Phi}$ is the set of transistors of doping $\Phi$ being part of the voltage biases in the circuit. $T_{cb, \Phi}$ is the set of transistors of the same doping $\Phi$ being part of current biases.

\subsection{Hierarchy Level 3:  Amplification Stage Subblocks}
The DC current flow must be symmetric in the subblocks of a non-inverting stage. We therefore define that the transconductor of the non-inverting stage $tc_{ninv,k}$ and its load $l_k$  must have symmetrical geometries:
\begin{equation}\label{eq:symDp}
\begin{split}
\forall_{tc_{ninv,k} \in \mathcal{M}_{tc_{ninv}}} & \{t_{k,i, \Phi}, t_{k,j, \Phi}\} = tc_{ninv,k} ~\\
\wedge~& l_{t_{k,i}} = l_{t_{k,j}} ~\wedge~ w_{t_{k,i}} = w_{t_{k,j}} 
\end{split}
\end{equation}
\begin{equation}\label{eq:symLoad}
\forall_{t_i,t_j \in ({T}_{l})} t_i.g \leftrightarrow t_j.g \Rightarrow (l_{t_i} = l_{t_j} \wedge w_{t_i} = w_{t_j})
\end{equation}
$\mathcal{M}_{tc_{ninv}}$ is the set of non-inverting transconductor in the op-amp and $T_l$ the set of transistors forming the load.

\subsection{Hierarchy Level 4: Op Amp Subblocks}
Symmetrical op-amps and fully differential two-stage op-amps have two second stages. They must be symmetrical:
\begin{equation} \label{eq;symSecondStage}
	\begin{split}
	\forall_{t_m \in a_{2,1}, t_n \in a_{2,2} } ~  &
	[(t_m.pos = t_n. pos) \\
 &	\rightarrow (w_{t_m} = w_{t_n} \wedge l_{t_m} = l_{t_n}) ]
	\end{split}
\end{equation}
$t_k.pos$ gives the position of a transistor, e.g., n-type transistor and connected to the ground net. The transistors on equal positions should have equal  geometries. In the symmetrical op-amp in Fig. \ref{fig:symmetricalOpAmp}, the transistors $N_4, N_5$ have equal positions and therefore should have the same  sizes. 
The other transistor pairs are $P_5, P_6$ and $P_3, P_4$.

\section{Functional Block Behavioral Constraints}\label{sec:finctionalBlocksPerformanceConstraints}

Behavioral constraints on a functional block are constraints on its transistor variables required to ensure the proper functionality of the block. Behavioral constraints are derived for structures (HL 2),  amplification stage subblocks (HL 3) and op-amp subblocks (HL 4).

\subsection{Hierarchy Level 2: Structures}
Behavioral constraints for specific types of current mirrors are instantiated on this level. An example is the cascode current mirror (e.g. Fig~\ref{fig:TelescopicOpAmp}, $N_1-N_4$). In this type of current mirror, the voltage potentials of the inner nets, e.g., Fig. \ref{fig:TelescopicOpAmp}, $n_6,n_7$, must be equal to suppress the effect of the channel length modulation.
To obtain equal voltages, the ratio of the widths of the transistors in the current mirror must be restricted:
\begin{equation}\label{eq:ccm}
\frac{w_{ccm,vb,d}}{w_{ccm,cb,d}} = \frac{w_{cm,vb,s}}{w_{cm,cb,s}}
\end{equation}
$w_{ccm,vb,d}, w_{ccm,vb,s}$ are the drain and the source transistor of the voltage bias in the cascode current mirror.
$w_{ccm,cb,d}, w_{ccm,cb,s}$ are the drain and the source transistor of the current bias in the cascode current mirror.
Please note that the  transistor length is already restricted by \eqref{eq:symVbCb}.

\subsection{Hierarchy Level 3:  Amplification Stage Subblocks}
A behavioral constraint on the amplification stage subblock level exists for the complementary transconductor $tc_c$.
The transconductance of the transistors in the n-doped differential pair $gm_{dp,{n,i}}|_{i=1,2}$ and p-doped differential pair $gm_{dp,{p,j}}|_{j=1,2}$ of  $tc_c$ must be equal. 
\begin{equation}
gm_{dp,n,i}|_{i=1,2} = gm_{dp,{p,j}}|_{j=1,2}
\end{equation}
Also the currents of the differential pairs generated with the n- and p-doped  transistors in the stage bias  $b_{s,c}$ must be equal:
\begin{equation}
|i_{DS,{b_{s,c,n}}}|= |i_{DS,b_{s,c,p,}}|
\end{equation}

\subsection{Hierarchy Level 4:  Op-Amp Subblocks}
A constraint on the op-amp subblock level is the output voltage offset constraint for two stage op-amps.
To suppress an offset voltage on the output voltage by equal input voltage, the voltage potentials at the first stage output nets $n_{a_1.out_1}, n_{a_1.out_2}$, e.g., Fig. \ref{fig:TelescopicOpAmp}, $n_{a_1.out_1} = n_5, n_{a_1.out_2} = n_8$ must be equal.
\begin{equation}\label{eq:OutputVoltageOffsetConstraint}
a_2 \in \mathcal{M} \Rightarrow	v_{n_{a_1.out_1}} = v_{n_{a_1.out_2}}
\end{equation}
$\mathcal{M}$ is the set of functional blocks of an op-amp topology.

\section{Intermediate Performance Equations} \label{sec:IntermediatePerformanceEquations}
Hierarchy levels 1, 3-5 are considered to establish all intermediate performance equations.
 They are only instantiated for a functional block if an op-amp performance equation requires them.

\subsection{Hierarchy Level 1: Devices}
The saturation drain-source voltage of a transistor and the net capacitance of a net in the circuit are equations generated based on the device information.

\subsubsection{Saturation Drain-Source Voltage}\label{sec:SaturationDrainSourceVoltage}
The saturation drain-source voltage is the voltage at least needed to keep a transistor in saturation. 
According to \eqref{eq:voltageConstraints}, this is:
\begin{equation}\label{eq:vdssat}
	v_{DS,sat, i} = \begin{cases}
		v_{GS,i}-v_{th,i}, &t_i.type = {nt}\\
		v_{GS,i}, &t_i.type = {dt}
	\end{cases}
\end{equation}
considering  that for a  diode transistor $dt_k$, $v_{GS,k} = v_{DS,k}$.

\subsubsection{Net Capacitance}\label{sec:NetCapacitance}
The  capacitance $C_{n_i}$ of a net $n_i$ depends on the pins $P_{{n_i}}$ connected to $n_i$:
\begin{equation}\label{eq:netCapacitance}
C_{n_i} = \sum_{p_j \in P_{{n_i}}}C_{p_j}
\end{equation}
$C_{p_j}$ is the capacitance arising by the pin $p_j$. The corresponding equations to calculate $C_{p_j}$  are given in \cite{MOSCapacitances}.

\subsection{Hierarchy Level 3: Amplification Stage Subblocks}
Transconductance and output conductances of the functional blocks are important properties to be described on this level.

\subsubsection{Transconductance}\label{sec:inputConductance}
The transconductance of a transconductor is defined by one of the transistors $t_{tc, i, in}$ whose gate is connected to the input signal of the stage.
\begin{equation}\label{eq:inputConductance}
	gin_{tc, i} = gm_{t_{tc,i, in}}
\end{equation} 
In a non-inverting stage, $t_{tc, i, in}$ is one of the transistors of the differential pair. Due to symmetry, the transconductance of both transistors is equal. In an inverting stage, $t_{tc, i, in}$ is the transistor whose gate is connected to the output of the previous stage. In op-amps with two second stages, the transconductance of only one of the two stages must be calculated due to symmetry.

For the calculation of the transconductance of the complementary transconductor (Fig. \ref{fig:railToRailAmplifier}),  the transconductances of the transistors in the pmos differential pair and the nmos differential pair must be considered:
\begin{equation}
gin_{tc_c} = gm_{t_{tc, in, n}} + gm_{t_{tc, in, p}}
\end{equation} 

\subsubsection{Output conductance}\label{sec:outputConductance}
The computation of $gout_i$ for a functional block $m_i$ depends on its inner structure. It is distinguished between functional blocks consisting of one- and two-transistor stacks. A transistor stack is defined as a sequence of transistors having a drain-source connection \cite{ABNG20}. $N_1,N_3$ in Fig. \ref{fig:TelescopicOpAmp} is an example of a two-transistor stack.
\begin{equation}\label{eq:outputConductanceTS}
gout_i =\begin{cases}
gd_{t_{i,out}},~~~~~~~~~~~~~~  \{t_{i,out} \} = ts_i \subseteq m_i \\
\frac{gd_{t_{i,out}}gd_{t_{i,supply}}}{gm_{t_{i,out}}}, ~~~ \{t_{i,out}, t_{i,supply} \} = ts_i \subseteq m_i 
\end{cases}
\end{equation}
If the functional block $m_i$ consists of one-transistor stacks, only the output conductance of the transistor in $m_i$ connected to the stage output is relevant for the calculation. If $m_i$ consists of two-transistor stacks, the transistor connected to the stage output and the transistor connected to the supply voltage rail are relevant. The type of functional block as load, transconductor and stage bias is irrelevant. All three types consist of one or two transistor stacks. If the functional block consists of two transistor stacks, the stacks are symmetrical. In non-symmetrical load parts \cite{ABNG20}, the output-connected transistor stack is only relevant for calculations. Hence, in Fig.~\ref{fig:symmetricalOpAmp},  $gout_{tc_1} = gd_{N_1} = gd_{N_2}$  and $gout_{tc_{2,2}} = \frac{gd_{P_5} gd_{P_3}}{gm_{P_5}}$.

Further differentiation must be made for load parts including a gate-connected couple being part of a cascode or folded-cascode differential pair (Sec. \ref{sec:functionalBlocks}):
\begin{equation}\label{eq:outputConductanceGCC}
gout_i =\begin{cases}
\frac{gd_{t_{i,out}}gd_{t_{tc_1,1}}}{gm_{t_{i,out}}}, ~~~~~~~ \{t_{i,out}\} = ts_i \subseteq m_i 
\\ ~~~~~~~~~~~~~~~~~~~~~~~~~~~~\wedge	 \{t_{tc_1,1}, t_{i,out}\} \subset cdp_k \\
\frac{gd_{t_{i,out}}(gd_{t_{i,supply}}  +gd_{t_{tc_1,1}})}{gm_{t_{i,out}}}, ~~~ \\
~~~~~~~~~~~~~~~~~~~~~~~~~\{t_{i,out}, t_{i,supply} \} = ts_i \subseteq m_i \\
~~~~~~~~~~~~~~~~~~~~~~~~~~~~\wedge	 \{t_{tc_1,1}, t_{i,out}\} \subset fcdp_k	\\
\end{cases}
\end{equation}
The output conductance of one of the transistors of the differential pair must be included in these calculations. Thus for the load part formed by $P_3, P_4$ in Fig.~\ref{fig:TelescopicOpAmp} $gout_{l_{p,1,1}} = \frac{gd_{P_4} gd_{P_2}}{gm_{P_4}}$
and in Fig.~\ref{fig:foldedCascodeOpAmp} $gout_{l_{p,1,1}} = \frac{gd_{P_2} (gd_{P_4} + gd_{N_2})}{gm_{P_2}}$.

In symmetrical op-amp and CMFB stages, the output conductance $gout_i$ of the load part is the transconductance of one of the transistors connected with its gate to the output of the stage. The load part consists only of voltage biases \cite{ABNG20}:
\begin{equation}\label{eq:outputConductanceVB}
gout_i =
gm_{t_{g.out}}, ~m_i.type = l_p \wedge m_i =\{vb_{i,1}, vb_{i,2}\}
\end{equation}
Hence, in Fig. \ref{fig:symmetricalOpAmp}, $gout_{l_{p,1,1}} = gm_{P_1} = gm_{P_2}$

\subsection{Hierarchy Level 4: Op-Amp Subblocks}
The output resistance, the open-loop gain and  the non-dominant poles and zeros of an amplification stage are calculated on this level. 

\subsubsection{Stage Output Resistance} \label{sec:outputRestistance}
The output resistance of an amplification stage is described by the output conductances of the $k$ functional blocks of the stage, e.g. stage biases, load parts and transconductors, connected to the stage output net. 
\begin{equation}\label{eq:outputRestistance}
R_{out,i} =	\frac{1}{\sum_{j=1}^{k} gout_i}
\end{equation}
For the symmetrical op-amp (Fig. \ref{fig:symmetricalOpAmp}), the output resistance of the first stage $R_{out, a_1}$ is for example calculated by $gout_{tc_1}$ and $gout_{l_{p,1,1}}$.
For the folded-cascode first stage in Fig. \ref{fig:foldedCascodeOpAmp}, the output resistance is
$R_{out,a_1} = \frac{1}{gout_{l_{p,1,1}} + gout_{l_{p,1,2}}}$.

\subsubsection{Stage Open-Loop Gain} \label{sec:stageOpenLoopGain}
The open-loop gain of a stage  $A_{D0,i}$ is calculated by the transconductance $gin_{tc,i}$ of its transconductors  (Sec. \ref{sec:inputConductance}) and its output resistance $R_{out, i}$:
\begin{equation}\label{eq:stageOpenLoopGain}
A_{D0,i} = gin_{tc,i} \cdot R_{out, i}
\end{equation}

\subsubsection{Stage Non-Dominant Poles}
Non-dominant poles arise for every stage in the op-amp. 
They must be calculated for every transistor $t_k$ on the signal path from input to output. The pole of the transistor $t_k$ is calculated by:
\begin{equation}
f_{ndp,t_k} = \frac{gm_{t_k}}{2\pi C_{n_j}}
\end{equation} 
$C_{n_j}$ is the net capacitance of the net the signal passes by before encountering $t_k$. 
It is calculated by \eqref{eq:netCapacitance} and contains the parasitics emerging from the transistor pins as well as the capacitance of the capacitors connected to the net.

If a compensation capacitor is connected between the input and the output of a stage, the equation of  non-dominant pole at the input transistor of the stages changes to:
\begin{equation}
	f_{ndp,{inv,C_c}} = \frac{gin_{tc_{inv}}}{2\pi (C_{n_{out}} + \frac{C_{n_{tc,{inv,in,g}}} \cdot C_{n_{out}}}{C_C}+ C_{n_{tc,{inv,in},g}})}
\end{equation}
$gin_{tc_{inv}}$	is the transconductance of the stage. $C_{n_{tc,{inv,in,g}}}$ is the capacitance of the gate net carrying the input signal of the stage. $C_{n_{out}}$ is the capacitance of the output net of the stage and $C_C$ the capacitance value of the compensation capacitor.

\subsubsection{Stage zeros}
Zeros  are evoked by non-dominant poles if they are a mirror pole, i.e., only half of the signal is influenced by it. They are set to occur at twice of the frequency of the mirror pole $f_{ndp, mir}$:
\begin{equation}
	f_{z, mir} = 2 \cdot f_{ndp, mir}
\end{equation}

\subsection{Hierarchy Level 5: Op-Amp}
The complete composition of the op-amp must be considered to calculate the dominant pole and the positive zero.

\subsubsection{Dominant Pole}
The dominant pole is the pole at the smallest frequency in an op-amp. It occurs mostly at the output net of the first stage and is calculated by:
\begin{equation}
f_{dp} = \frac{1}{2 \pi C_{n_{out}} gin_{tc_2} \Pi_{i=1}^2 R_{out,i}}
\end{equation}
$C_{n_{out}}$ is the capacitance at the output net of the first stage, $gin_{tc_2}$ the transconductance of the  second stage transconductor and $R_{out,j}$ the output resistance of the amplification stages. For single-stage op-amps, $gin_{tc_2}$ and $R_{out,2}$ are set to one.

In symmetrical op-amps, the dominant pole occurs at the output net of the second stage. The equation changes to:
\begin{equation}
	f_{dp} = \frac{1}{2 \pi C_{n_{out}} gm_{tc_3} \Pi_{i=2}^3 R_{out,i}}
\end{equation}
where $C_{n_{out}}$ is the capacitance at the output net of the second stage. If no third stage is part of the symmetrical op-amp, $gm_{tc_3}$ and $R_{out,3}$ are set to one.

\subsubsection{Positive Zero}
In op-amps with compensation capacitor $c_C$, a positive zero exists.
It is calculated by:
\begin{equation}
f_{pz} = \frac{1}{2\pi C_C (\frac{1}{gin_{tc_{inv,k}}} - \frac{1}{gd_{R_C}})}
\end{equation}
$gin_{tc_{inv,k}}$ is the transconductance of the inverting transconductor  connected by $c_C$ to a previous stage. If a compensation resistor $R_C$ is part of the circuit, $gd_{R_C}$ 
is the output conductance of the transistor emulating the compensation resistor,  otherwise $gd_{R_C} = 1$.

\section{Op-Amp Performance Equations}\label{sec:opAmpPerformanceConstraints}

Analogous to the  equations and constraints before, the performance features of an op-amp are ordered hierarchically. Some performance equations only need the device level information as input (HL 1). Others are based on op-amp subblocks (HL~4) or on the whole op-amp (HL 5).

\subsection{Hierarchy Level 1: Devices}
The area and quiescent power of the op-amp is calculated
based on device level information.

\subsubsection{Area}\label{sec:Area}
An estimation of the area of the circuit is calculated through the gate areas of all $k$  transistors:
\begin{equation}\label{eq:ConstraintsArea}
z_D = \sum_{i=1}^k W_i \cdot L_i
\end{equation}

\subsubsection{Quiescent Power}\label{sec:PowerConsumption}

The quiescent power $z_{QP}$ is the product of the positive supply voltage $v_{VDD}$ subtracted by negative voltage $v_{VSS}$ with the sum of the $n$ currents flowing into the positive supply voltage net $n_{VDD}$. If the bias current of the circuit $i_{Bias}$ is applied to an nmos transistor, it must be added to the currents flowing into $n_{VDD}$.
\begin{equation}\label{eq:ConstraintsPower}
z_{QP} = (v_{VDD} - v_{VSS}) \cdot \begin{cases} \sum_{j=1}^n |i_j|,  &t_{Bias}.\Phi =p\\  \sum_{j=1}^n |i_j| + i_{bias}, &t_{Bias}.\Phi =n
\end{cases}
\end{equation}

\subsection{Hierarchy Level 4: Op-Amp Subblocks}
Performance features determined by one amplification stage are formulated on this level.
These are common-mode input voltage, output voltage, common-mode rejection ratio (CMRR) and unity-gain bandwidth.

\subsubsection{Common-mode Input Voltage}\label{sec:CommonModeInputVoltage}
The common-mode input voltage describes the range in which the input voltage can vary without changing the behavior of the op-amp. We can specify a maximum $z_{v_{cm,max}}$ and a minimum $z_{v_{cm,min}}$ common-mode input voltage.  
The two voltage loops which describe $z_{v_{cm,max}}, z_{v_{cm,min}}$
are either over the load of the first stage $l_{1}$ or over its  stage bias $b_{s,1}$. Therefore, we can define the two limiting values by $v_{cm,l_{1}}, v_{cm,b_{s,1}}$.

 $z_{v_{cm,max}}, z_{v_{cm,min}}$ are defined depending which of $v_{supply,b_{s,1}},v_{supply,l_{1}}$ equals $v_{VDD},v_{VSS}$.
\begin{equation}
\begin{split}
v_{supply,b_{s,1}} = v_{VDD} \wedge v_{supply,l_{1}} = v_{VSS} \\
\Rightarrow z_{v_{cm,max}} = v_{cm,b_{s,1}} \wedge z_{v_{cm,min}} = v_{cm,l_{1}} \\
v_{supply,b_{s,1}} = v_{VSS} \wedge v_{supply,l_{1}} = v_{VDD}
\\\Rightarrow z_{v_{cm,max}} = v_{cm,l_{1}} \wedge z_{v_{cm,min}} = v_{cm,b_{s,1}}
\end{split}
\end{equation}
For loads  connected to both supply voltage rails, e.g. Fig. \ref{fig:foldedCascodeOpAmp}, the supply voltage rail opposite to $v_{supply,b_{s,1}}$ is considered.

If the transistors in the paths are in saturation and in strong inversion, $v_{cm,b_{s,1}}$ and $v_{cm,l_{p,1,1}}$ are defined by the minimum/maximum voltage which is allowed when keeping all transistor in saturation.
For a single transistor, this voltage is defined by the minimum saturation voltage \eqref{eq:vdssat}.
Hence,  $v_{cm,b_{s,1}}$ is defined as:
\begin{equation}
	v_{cm,b_{s,1}} := v_{supply,b_{s,1}} + v_{GS,tc_1} + \sum_{i=1}^{|b_1|} v_{DS,sat,i}
\end{equation}

Determining $v_{cm,l_{1}}$ is more complex, as the structure of the load highly varies \cite{ABNG20}. Two relevant voltage paths $e_1, e_2$ exist having the smallest possible number of voltage drops. Each path starts from one of the outputs of the first stage transconductor $tc_1$ going to the supply-voltage rail of the load $n_{supply,l_1}$. In the folded-cascode op-amp (Fig. \ref{fig:foldedCascodeOpAmp}),  $e_1 =\{P_3\}, e_2 = \{P_4\}$. In the telescopic op-amp, (Fig. \ref{fig:TelescopicOpAmp}) $e_1 =\{P_3,N_1,N_3\}, e_2 = \{P_4, N_2,N_4\}$.
For the calculation of $v_{cm,l_1}$ the path $e_k$ is chosen with the most diode transistors. For diode transistors, $v_{DS, sat} = v_{GS}$, such that they have a much higher impact on the input voltage range as normal transistors. 
 If the transistors in the paths are in saturation and in strong inversion, $v_{cm,l_{1}}$ is defined by:
\begin{equation}
\begin{split}
v_{cm,l_{1}} := &~ v_{supply,l_1} + v_{th,tc_1}  \\
&+\sum_{m=1}^{|e|}\begin{cases}
-(v_{DS,sat, m}), &t_m \subset {gcc_m}\\
v_{DS,sat, m}, &else\\
\end{cases}
\end{split}
\end{equation}
$v_{th,tc_1}$ is the threshold voltage of a transistor of the transconductor of the first stage $tc_1$. 
If a transistor of a gate-connected couple $gcc_k$ is in the path $e$, e.g. $P_3,P_4$ in Fig. \ref{fig:TelescopicOpAmp}, it introduces a negative value for $v_{DS,sat,k}$. It has a different substrate doping than the other transistors in the load relevant for $e$. 

\subsubsection{Output Voltage}\label{sec:OutputVoltage}
The output voltage swing is described by the last stage of an op-amp. A maximum value $z_{v_{out,max}}$ and a minimum value $z_{v_{out,min}}$ are defined by the shortest paths from the output of the op-amp to the supply-rails $e_{VDD}, e_{VSS}$. The paths contain the transistors being part of transistor stacks connecting the supply-voltage rails to the output. If the transistors are supposed to be in saturation and in strong inversion, the corresponding equations are:
\begin{equation}
\begin{split}
z_{v_{out,max}} = v_{VDD} +\sum_{i=1}^{|e_{VDD}|} v_{DS, sat, i}
\end{split}
\end{equation}
\begin{equation}
\begin{split}
z_{v_{out,min}} = v_{VSS} +\sum_{i=1}^{|e_{VSS}|} v_{DS, sat, i}
\end{split}
\end{equation}
with $v_{DS, sat, i}$ described by \eqref{eq:vdssat}.

\subsubsection{Common-mode Rejection Ratio} \label{sec:CMRR}
For non-fully differential op-amp topologies, an analytical equation can be derived that gives a good approximation of the static systematic common-mode rejection ratio (CMRR$_s$). For all op-amps but symmetrical op-amps, the CMRR$_s$ only depends on the structure of the first stage.
\begin{equation}
	z_{CMRR} = 2A_{D0,1} \cdot \frac{gm_{l_1,g.out}}{gout_{b_{s,1}}}
\end{equation}
$A_{DO,1}$ is the open-loop gain of the first stage of the op-amp \eqref{eq:stageOpenLoopGain}. $gout_{b_{s,1}}$ the output conductance of the first stage bias calculated according to Sec. \ref{sec:outputConductance}. $gm_{l_1,g.out}$ is the transconductance of the load transistor connected with its gate to one of the output nets of the first stage. 
If the gates of two load transistors are connected to the output of the first stage, any of these two can be chosen. They have equal $gm$-values, as the load of an op-amp is  symmetric.
If no load transistor's gate is connected to the output of the first stage the equation to calculate the CMRR$_s$ changes:
\begin{equation}
z_{CMRR} = 2 \cdot \frac{gin_{tc_1}}{gout_{b_{s,1}}}
\end{equation}
$gin_{tc_1}$ is the transconductance of  the transconductor of the  first stage calculated according to Sec. \ref{sec:inputConductance}. 

The CMRR of the symmetrical op-amp is also defined by the open-loop gain of the second stage $A_{D0,2}$:
\begin{equation}
z_{CMRR_{sym}} = 2A_{D0,1} \cdot A_{D0,2} \cdot \frac{gm_{l_1,g.out}}{gout_{b_{s,1}}}
\end{equation}

For fully-differential op-amps, the CMRR$_s$ also depends on the common-mode feedback circuit and is not discussed in this paper. In complementary op-amps, the two stage bias types of the first stage, pmos and nmos, must be considered.

\subsubsection{Unity-gain bandwidth}\label{sec:UnityGainBandWidth}
The unity-gain bandwidth $z_{f_{GBW}}$ is calculated by the first stage transconductor  $tc_1$ and the capacitance of the first stage output net $C_{n_{a_1, out}}$:
\begin{equation}\label{eq:UnitiGanBandwidthNormal}
z_{f_{GBW}} = \frac{gin_{tc_1}}{2 \pi C_{n_{a_1, out}}}
\end{equation}

The equation for $z_{f_{GBW}}$ differs slightly for symmetrical op-amps, as the second stage impacts the unity-gain bandwidth:
\begin{equation}\label{eq:UnitiGanBandwidthSym}
z_{f_{GBW}} = \frac{A_{D0,1} \cdot gin_{tc_2}}{2 \pi C_{n_{a_2,out}}}
\end{equation}
$A_{D0,1}$ is the first stage open-loop gain, $gin_{tc_2,}$ the transconductance of the second stage transconductor, and $C_{n_{a_2,out}}$ the   capacitance  of the second stage output net connected to a capacitor.

\subsection{Hierarchy Level 5: Op-Amp}
The overall op-amp structure is considered for the calculation of open-loop gain, the slew rate and the phase margin.

\subsubsection{Open-loop Gain}\label{sec:OpenLoopGain}
The open-loop gain of an op-amp is the product of the open-loop gains of its $n$ stages:
\begin{equation}\label{eq:openLoopGain}
z_{A_{D0}} = \prod_{k=0}^{n} A_{D0,i}
\end{equation}

\subsubsection{Slew Rate}\label{sec:SlewRate}
The slew rate $z_{SR}$ of a circuit is calculated from the bias currents of the $n$ stages and the capacitances of the output nets of the stages,
\begin{equation}
z_{SR} = min\{\frac{|i_{DS,{b_{s,1}}}|}{C_{n_{out,1}}},.., \frac{|i_{DS,{b_{s,n}}}|}{C_{n_{out,n}}} \}
\end{equation}
where $i_{DS,{b_{s,k}}}$ is the drain-source current of a transistor part of the stage bias $b_{s,k}$ of the stage $k$.
$C_{n_{out,k}}$ is the capacitance of the stage output net calculated by \eqref{eq:netCapacitance}.
For symmetrical op-amps, the first stage output net does not have to be considered. However, if one of the input transistors of the first stage is shut down, the bias current of the first stage is amplified and mirrored by the current mirror forming the   first stage load and the second stage transconductor, e.g., Fig. \ref{fig:symmetricalOpAmp} $P_2, P_4$. Therefore,  twice the bias current of the second stage must be considered during slew rate calculations.

In a folded-cascode op-amp, the current $i_{DS, l_{B,GCC}}$ of the two transistors biasing the gate-connected couple, e.g., Fig. \ref{fig:foldedCascodeOpAmp} $P_3, P_4$, must be considered during slew rate calculation. The smallest current of $i_{DS, l_{B,GCC}}$,  $i_{DS,{b_{s,1}}}$ restricts the slew rate.

\subsubsection{Phase Margin}\label{sec:PhaseMargin}

The phase margin  $z_{PM}$ is calculated by the non-dominant poles and zeros of the circuit:
\begin{equation}\label{eq:phaseMargin}
z_{PM} = \frac{\pi}{2} - \sum_{i=1}^{m} \text{atan}(\frac{f_{GBW}}{f_{ndp_i}})  + \sum_{j=1}^{n} \text{atan}(\frac{f_{GBW}}{f_{z_j}})
\end{equation}
$f_{GBW}$ is the unity-gain bandwidth of the circuit. A positive zero has a negative influence on the phase margin, like non-dominant poles. 

The non-dominant poles and zeros must be at least an order of magnitude larger than the dominant pole.
\begin{equation}
\forall_{f_{i} \in (F_{ndp} \cup F_{z})}  \frac{f_i}{f_{dp}} >10
\end{equation}

\section{Automatic Instantiation of the Equation-based Circuit Model Based on HPEL} \label{sec:AutomaticModelCreation}
\begin{figure}\centering
	\includegraphics[width=0.75\linewidth]{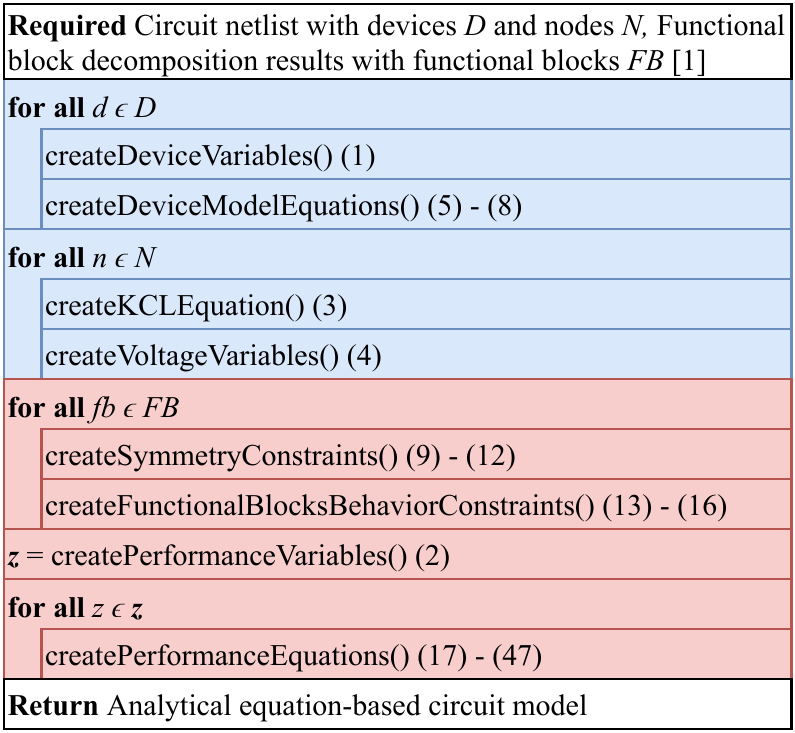}
	\caption{Automatic instantiation of an equation-based circuit model for a given topology}
	\label{fig:modelCreation}
\end{figure}

Fig. \ref{fig:modelCreation} shows the automatic synthesis of the equation-based circuit model for a given op-amp topology. The input of the algorithm are the circuit netlist and the results of the functional block decomposition method in \cite{ABNG20}, which automatically identifies all functional blocks described in Sec.  \ref{sec:functionalBlocks} in a circuit netlist. The basic circuit model and the circuit performance model are automatically instantiated based on this input.

The algorithm iterates over the devices and nodes in the circuit to create the basic model. The corresponding variables and equation are automatically instantiated. This is similar to a circuit simulation tool.

Symmetry constraints,  functional block behavior constraints and  performance equations are automatically created to form the op-amp performance model. The symmetry and functional block behavior constraints are created by iterating over all recognized functional blocks, instantiating the corresponding constraints by selecting the corresponding variables of the basic circuit model. This is similar to the method in~\cite{Massier2008}, which creates constraints for basic transistor pairs. 

\begin{figure}\centering
	\includegraphics[width=0.65\linewidth]{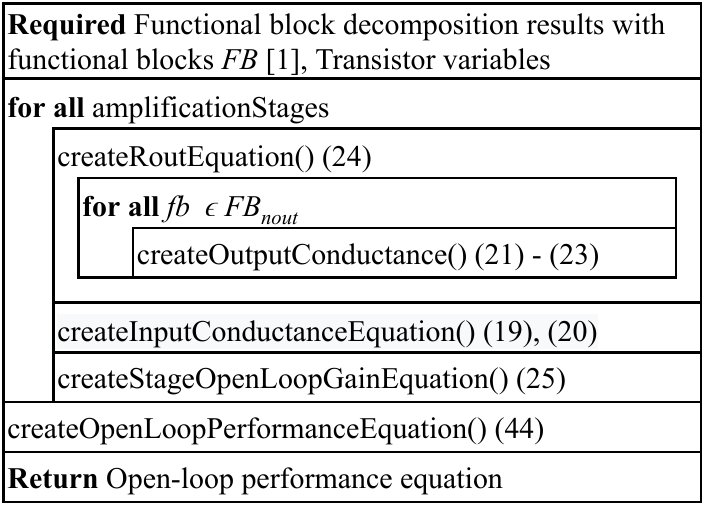}
	\caption{Performance equation creation on the example of the open loop gain}
	\label{fig:openLoopGainEquationCreation}
\end{figure}

The performance equations are set up for every performance variable in \eqref{eq:performanceVariable}. Every equation stated on a high level of abstraction in Sec. \ref{sec:opAmpPerformanceConstraints} is broken down into the circuit variables using the intermediate performance equations. Fig.~\ref{fig:openLoopGainEquationCreation} illustrates this procedure with the open-loop gain. To instantiate the open-loop gain performance equation~\eqref{eq:openLoopGain}, the  open-loop gain equations of the individual amplification stages must be created. These equations take the output resistances and the transconductances of the stages as input \eqref{eq:stageOpenLoopGain}. The transconductance of a stage in turn takes the circuit variables as input (Sec. \ref{sec:inputConductance}). For the equation of the output resistance, the equations of  the output conductances of all functional blocks on HL 3 $FB_{n_{out}}$  connected to the output net $n_{out}$  must be created. The equations of the output conductances  has the circuit variables as input (Sec. \ref{sec:outputConductance}). Thus, an overall open-loop gain  equation is automatically instantiated with the circuit variables as input.

Analogously to Fig.~\ref{fig:openLoopGainEquationCreation}, the performance equations for every supported performance feature 
in~\eqref{eq:performanceVariable} are automatically instantiated for a given topology linking the abstract performance equations in Sec.~\ref{sec:opAmpPerformanceConstraints} to the circuit variables using the intermediate performance equations (Sec.~\ref{sec:IntermediatePerformanceEquations}).
Many intermediate performance equations are part of several different op-amp performance equations. The transconductance of the first stage is for example part of the open-loop gain equation as well as part of the equation for the unity-gain bandwidth.  
The equations are stored topology-independent but customized by the algorithms.
In contrast to the presented approach, the state of the art is limited to individual topologies and their specific equation sets.

\section{Application of the Hierarchical Performance Equations Library in Automatic Sizing} \label{sec:InitialSizingTool}

The circuit model automatically created for a topology with the method in Sec.~\ref{sec:AutomaticModelCreation} can be applied to size the circuit for performance requirements given as lower and upper bounds and for a given process technology.
Intermediate performance requirements, e.g., on poles, are automatically derived from the given op-amp specification.

The automatically created circuit model is fed into a suitable solver. We use constraint programming~\cite{HBCP} in this work.  Constraint programming is suitable for the combinatorial character of analog circuit sizing due to the manufacturing-induced discrete value range of transistor geometries. It allows all function types, as e.g., trigonometrical, polynomial. No  further approximations must be made to the performance equations.
A detailed description and the adaptations we made to the constraint programming solver are given in \cite{ABNG20b}.
During the sizing process, the basic model of the circuit emulates the circuit simulation, while the performance model (Secs.~\ref{sec:symmetryConstraints} - \ref{sec:opAmpPerformanceConstraints}) describes the overall behavior of the functional blocks of the given op-amp topology providing the information needed for transistor sizing.

The tool needs a few seconds to find the first initial sizing for a circuit. The results are further improved towards higher performance safety margins
by letting the optimizer run for one more minute. After one minute, the improvement slowed down significantly in the experiments, therefore the optimization loop has been set to run for one minute overall.

The runtime equals the runtime of numerical sizing methods, e.g., \cite{MARS}, which also have small runtimes on modern hardware due to parallelized processes.
However, the lower computational cost of this method can be demonstrated by integrating the method into a synthesis tool featuring thousands of different circuits \cite{SMACD2021}. In this context, the method is twice as fast as state-of-the-art numerical approaches, e.g., \cite{FEATS}.

Note that for the numerical optimization techniques, the constraints, parameters, performance features, simulation configuration, and waveform postprocessing,  must be  set up for every circuit before starting the optimization.

\section{Experimental Results}\label{sec:ExperimentalResults}

This section presents experimental results for the four circuits in Fig. \ref{fig:differentOpAmpToplogies}. We present the performance models automatically generated  with the algorithm in Sec. \ref{sec:AutomaticModelCreation} as well as
transistor dimensions (Table~\ref{tab:dimensionsOpAmp}) and performance values obtained with the circuit models (Tables~\ref{tab:GenericSpecs} and \ref{tab:OtherSpecs}).

\subsection{Performance model}

In the following, the important parts of the performance models of the four circuits in Fig. \ref{fig:differentOpAmpToplogies} are described. 
All equations were generated individually and automatically using the algorithms in Sec. \ref{sec:AutomaticModelCreation}. 
The generated circuit models correspond well to the models presented in analog design books \cite{Allen,AnalogIntegratedCircuitDesign,LakerSansen, MOSCapacitances}.

\subsubsection{Symmetry Constraints}
Table \ref{tab:symmetryConstraints} shows the symmetry constraints derived for the four circuits in Fig. \ref{fig:differentOpAmpToplogies}. Eight symmetry constraints for basic structures were derived for the telescopic op-amp. This is identical to the number of current biases in the circuit. 
The large number of symmetry constraints for the amplification stage subblock level in the folded-cascode op-amp with CMFB results from the common-mode feedback (CMFB) stage in which both differential pairs must be identical.
As the two second stages $a_{2,1}, a_{2,2}$ in the symmetrical op-amp with high PSRR  must be symmetric, four symmetry constraint were derived for HL 4 for this circuit. 

\begin{table}\centering\scriptsize
	\setlength{\tabcolsep}{0.1cm}
	\caption{Symmetry constraints}\label{tab:symmetryConstraints}	
	\begin{tabular}{|>{\RaggedRight\arraybackslash}m{1.8cm}|>{\centering\arraybackslash}m{1.0cm}|>{\centering\arraybackslash}m{1.3cm}|>{\centering\arraybackslash}m{1.7cm}|>{\centering\arraybackslash}m{1.7cm}|}
		\hline
		& Telescopic op-amp& Symmetrical op-amp  & Folded-cascode op-amp  & Comp\-le\-men\-tary op-amp \\\hline
		HL 2: Structures & 8& 5  & 7& 7\\\hline
		HL 3:  Ampli\-fi\-ca\-tion stage subblocks & 5 &  3  & 12 & 8 \\\hline
		HL 4: Op-amp subblocks & - & 4 & - & -\\\hline
	\end{tabular}
\end{table} 

\subsubsection{Functional Block Constraints}
As a cascode current mirror forms one load part of the telescopic op-amp (Fig.~\ref{fig:TelescopicOpAmp}), its widths are restricted by the corresponding behavioral constraint \eqref{eq:ccm}. Furthermore, as the telescopic op-amp has a second stage, the output voltages of the first stage, i. e., the voltage potentials of the nets $n_5$,$n_8$, must be equal  \eqref{eq:OutputVoltageOffsetConstraint}. 

To  make the combination of folded-cascode op-amp and CMFB circuit work, a functional block constraint on the fifth hierarchy level not mentioned before must be generated for the folded-cascode op-amp with CMFB (Fig. \ref{fig:foldedCascodeOpAmp}). The unity-gain bandwidth of the CMFB circuit must be greater than the one of the op-amp, such that the CMFB circuit is faster.
\begin{equation}
f_{GBW, CMFB} > f_{GBW, op-amp}
\end{equation}
The unity-gain bandwidth is calculated according to \eqref{eq:UnitiGanBandwidthNormal} treating the CMFB stage as a first stage.

As the complementary op-amp (Fig. \ref{fig:railToRailAmplifier}) has a complementary first stage, the functional block constraints for HL~3 are generated restricting the first stage transconductor to have equal transconductances, and the stages biases to produce equal currents.

\subsubsection{Performance Equations}
In the following, the performance equations of the four circuit in Fig. \ref{fig:differentOpAmpToplogies} are presented.  We focus on the differences between the four circuits.

{\em Quiescent power:} 
For the telescopic op-amp, the currents following into the  positive supply voltage rail are considered to calculate the power consumption, while for the other three circuits also the bias currents of the circuit must be considered as it is applied to nmos transistors.
The equation for the quiescent power of the telescopic op-amp is:
\begin{equation}
\begin{split}
	z_{QP} = &(v_{VDD} - v_{VSS})  \\
&\cdot	(|i_{DS, P_7}| + |i_{DS, P_9}| + |i_{DS, P_5}| + |i_{DS, P_8}|)
\end{split}	
\end{equation}
and for the quiescent power of the symmetrical op-amp with high PSRR:
\begin{equation}
\begin{split}
z_{QP} = &(v_{VDD} - v_{VSS})  \cdot (|i_{DS, P_3}| + |i_{DS, P_1}|  \\
&+ |i_{DS, P_2}| + |i_{DS, P_4}| + |i_{DS, P_7}| + |i_{DS, N_7}|)
\end{split}
\end{equation}

{\em Common-mode input voltage:}
For the telescopic op-amp, the maximum input voltage is set by the path over the first stage stage bias:
\begin{equation}\label{eq:telescopicOpAmpVCMBIAS}
	z_{v_{cm,max}} = v_{cm,b_{s,1}} = v_{VDD}  + v_{GS,P_1} +  v_{GS,P_5}  - v_{th,p}
\end{equation}

For the symmetrical op-amp with PSRR and the folded-cascode op-amp, $z_{v_{cm,max}}$ is set by the path over the load.
For the  folded-cascode op-amp with CMFB, this is:
\begin{equation}
\begin{split}
z_{v_{cm,max}} =  ~v_{cm,l_{1}} = ~ v_{VDD} + v_{th,n}  + v_{GS, P_3} - v_{th,p} 
\end{split}
\end{equation}

In the telescopic op-amp, the load  defines the minimum input voltage. As higher minimum saturation voltages  must be respected, the load path with the two diode transistors $N_1, N_2$ is selected:
\begin{equation}
\begin{split}
z_{v_{cm,min}}	= &~ v_{ss} + v_{th,p} - (v_{GS, P_3} - v_{th,n}) + v_{GS, N_1} + v_{GS,N_2}
\end{split}
\end{equation}

In the other circuits, the minimum input voltage is restricted by the stage bias of the first stage, which leads to similar equations as in  \eqref{eq:telescopicOpAmpVCMBIAS} with the negative supply voltage as input. No equations are generated for the complementary op-amp, as it is assumed to allow all values as input voltage.

{\em Output voltage:}
In the telescopic op-amp and the symmetrical op-amp with high PSRR, the output voltage is restricted by one transistor on each path of the output stage. The output voltage equations for the telescopic op-amp are:
\begin{equation}
\begin{split}
z_{v_{out,max}} = & ~v_{VDD} + v_{GS, P_6} - v_{th,p} \\
z_{v_{out,min}} = &~v_{VSS} + v_{GS, N_5} - v_{th,n}
\end{split}
\end{equation}
For the folded-casode op-amp with CMFB and the complementary op-amp,
the output voltage is restricted by the load parts of the first stage, as the first stage is also the output stage. For each path, two transistors must be considered. For the folded-cacode op-amp, the output voltage is  restricted by:
\begin{equation}
\begin{split}
z_{v_{out,max}} = &~v_{VDD} + v_{GS, P_4} - v_{th,p} + v_{GS, P_2} - v_{th,p} \\
z_{v_{out,min}} = & ~v_{VSS} +v_{GS, N_8} - v_{th,n} + v_{GS, N_6} - v_{th,n}
\end{split}
\end{equation}

{\em Common-mode rejection ratio:}
The common-mode rejection ratio is calculated for the telescopic op-amp and the symmetrical op-amp with high PSRR.
To calculate CMRR of the telescopic op-amp, the open-loop gain of the first stage is needed. Using~\eqref{eq:stageOpenLoopGain}, we obtain: \eqref{eq:outputConductanceTS}, \eqref{eq:inputConductance}:
\begin{equation}\label{eq:telescopicOpAmpFirstStageGain}
	A_{D0,1} = \frac{gm_{P_1}}{\frac{gd_{P_1} \cdot gd_{P_3}}{gm_{P_3}} + \frac{gd_{N_1} \cdot gd_{N_3}}{gm_{N_1}}}
\end{equation}
$N_1$ is the load transistor chosen for the CMRR calculation as its gate is connected to an output of the first stage, which leads to following CMRR equation of the telescopic op-amp:
\begin{equation}
	z_{CMRR} = 2A_{D0,1} \cdot \frac{gm_{N_1}}{gd_{P_5}}
\end{equation}
As stated in Sec. \ref{sec:CMRR}, the first and second stage gain must be considered to calculate CMRR for symmetrical op-amps. For the symmetrical op-amp with high PSRR, these are:
\begin{equation}\label{eq:symmetricalOpAmpGain}
A_{D0,1} =  \frac{gm_{N_1}}{gd_{N_1} + gm_{P_1}}, ~~ A_{D0,2} = \frac{gm_{P_4}}{\frac{gd_{P_6} + gd_{P_4}}{gm_{P_6}} + gd_{N_5} }
\end{equation}
The equation of CMRR then is:
\begin{equation}
z_{CMRR} = 2A_{D0,1} \cdot  A_{D0,2}\frac{gm_{P_2}}{gd_{N_3}}
\end{equation}

{\em Unity-gain bandwidth:}
The unity-gain bandwidth is calculated similarly for the telescopic op-amp, for the folded-cascode op-amp with CMFB and for the complementary op-amp.
For the telescopic op-amp, it is:
\begin{equation}\label{eq:telescopicOpAmpGBW}
z_{f_{GBW}} = \frac{gm_{P_1}}{2 \pi C_{n_8}}
\end{equation}
In the complementary op-amp, both nmos and pmos differential pairs must be considered  to calculate the transconductance of the first stage transconductor (Sec.~ \ref{sec:inputConductance}). 

In the symmetrical op-amp, also the second stage must be considered to calculate the unity-gain bandwidth \eqref{eq:UnitiGanBandwidthSym}:
\begin{equation} 
z_{f_{GBW}} = \frac{A_{D0,1} \cdot gm_{P_4}}{2 \pi C_{n_5}}
\end{equation}

{\em Open-loop gain:}
The open-loop gain is calculated by the multiplication of the gain of the stages. 
Two stages must be considered in the telescopic op-amp (Fig \ref{fig:TelescopicOpAmp}), three stages in the symmetrical op-amp (Fig \ref{fig:symmetricalOpAmp}).
As the folded-cascode op-amp with CMFB consists of one stage only, its open-loop gain is the gain of the first stage.

In the complementary op-amp, two gate-connected couples exist.  The open-loop gain is therefore calculated by:
\begin{equation} 
z_{A_{D0}} =  \frac{ gm_{N_4} + gm_{P_4}}{\frac{gd_{P_8} \cdot (gd_{P_6} + gd_{N_4})}{gm_{P_8}} + \frac{gd_{N_8} \cdot (gd_{N_6} + gd_{P_4})}{gm_{N_8}}}
\end{equation}

{\em Slew rate:}
In the telescopic op-amp, the first stage and the second stage bias current must be considered for the slew rate:
\begin{equation}
z_{SR} = min\{\frac{|i_{DS,P_5}|}{C_{n_8}}, \frac{|i_{DS,{P_6}}|}{C_{n_{out}}} \}
\end{equation}
Please note that the capacitance of net $n_{8}$ is mainly influenced by the compensation capacitor $c_C$, the capacitance of net $n_{out}$  by the load capacitor $c_L$.

In the symmetrical op-amp, the second stage and the third stage  are considered for slew rate calculation as stated in Sec.~\ref{sec:SlewRate}. Twice the current of the second stage is considered.
\begin{equation}
z_{SR} = min\{\frac{2 \cdot |i_{DS,N_5}|}{C_{n_5}}, \frac{|i_{DS,{N_6}}|}{C_{n_{out}}} \}
\end{equation}

In the folded-cascode op-amp with CMFB, the bias currents of the first stage differential pair as well as the gate connected couple must be considered. This leads to:
\begin{equation}
z_{SR} = \frac{min\{|i_{DS, N_4}|, |i_{DS, P_4}|\}}{C_{n_{out}}}
\end{equation}
The same considerations must be made for the complementary op-amp. In addition, the pmos and nmos stage biases are of interest:
\begin{equation}
z_{SR} = \frac{min\{(|i_{DS,N_2}| + |i_{DS,P_2}|), (|i_{DS,N_6}| + |i_{DS,P_6}|)\}} {C_{n_{out}}} 
\end{equation}

{\em Phase margin:}
Two non-dominant poles are identified for the telescopic op-amp: one pole for the first stage and one for the second stage. The compensation capacitor brings a positive zero along. Hence, the automatically generated equation for the phase margin is:
\begin{equation}\label{eq:telescopicOpAmpphaseMargin}
z_{PM} = \frac{\pi}{2} - \text{atan}(\frac{f_{GBW}}{f_{ndp,a_1}}) - \text{atan}(\frac{f_{GBW}}{f_{ndp,a_2}}) - \text{atan}(\frac{f_{GBW}}{f_{pz}})
\end{equation}

In the symmetrical op-amp, three non-dominant poles arise: the first stage non-dominant pole, the non-dominant pole evoked by the compensation capacitor in the third stage and the non-dominant pole of the cascode transconductors in the second stages. The compensation capacitor  also leads to a positive zero. The  equation for the phase margin is:
\begin{equation}
\begin{split}
z_{PM} = \frac{\pi}{2} - & \text{atan}(\frac{f_{GBW}}{f_{ndp,a_1}}) - \text{atan}(\frac{f_{GBW}}{f_{ndp,a_3}}) \\
&- \text{atan}(\frac{f_{GBW}}{f_{ndp,a_{c,2}}}) - \text{atan}(\frac{f_{GBW}}{f_{pz}})
\end{split}
\end{equation}

For the folded-cascode op-amp with CMFB, the phase margins of the first stage and the CMFB circuit must be calculated. As the phase margin of the CMFB circuit is restricted by the non-dominant poles of the first stage and the CMFB stage, this phase margin is the most restrictive one.

In the complementary op-amp, two non-dominant poles of the first stage must be calculated, respecting the two differential pairs.

\subsection{Sizing Results}\label{sec:telescopicOpAmpSizingResults}
\begin{table}[tbp]\scriptsize\centering
		\caption{Dimensions for the circuits in Fig. \ref{fig:differentOpAmpToplogies}}\label{tab:dimensionsOpAmp}
	\subfloat[Telescopic op-amp]{
		\label{tab:DimensionsTelescopicOpAmp}
		\tabulinesep=0.3mm
		\begin{tabu}{|>{\RaggedRight\arraybackslash}p{2.3cm}|r|}
			\hline
			Variable & Value  \\\hline \hline
			$W_{P_1}= W_{P_2}$ &	172$\mu$m\\
			$W_{P_3}= W_{P_4} $ &	27$\mu$m\\
			$W_{P_5}$ &	247$\mu$m\\
			$W_{P_6}$ &	515$\mu$m\\
			$W_{P_7}$ &	7$\mu$m\\
			$W_{P_8}	$& 7$\mu$m\\
			$W_{P_9}$ &	43$\mu$m\\
			$W_{N_1}=W_{N_2}$ &	90$\mu$m\\
			$W_{N_3}=W_{N_4}$&	90$\mu$m\\
			$W_{N_5}$ &	130$\mu$m\\
			$W_{N_6}$&	269$\mu$m\\
			$W_{N_7} $&	166\\$L_{P_1} = L_{P_2}$ &	9$\mu$m \\
			$L_{P_3}=L_{P_4}=L_{P_8}$ &	4$\mu$m\\
			$l_{P_5} = L_{P_6} = L_{P_7}=L_{P_9} $&	3$\mu$m\\
			$L_{N_1}=L_{N_2}$ &	1$\mu$m\\
			$L_{N_3}=L_{N_4}$	&1$\mu$m\\
			$L_{N_5}$ &	1$\mu$m\\
			$L_{N_6}=L_{N_7}$ &	9$\mu$m\\
			$C_c$ &	6.4pF\\
			\hline
		\end{tabu}
	}%
	\qquad
	\subfloat[Folded-cascode op-amp with CMFB]{
		\label{tab:DimmensionFoldedCascodeOpAmp}
		\tabulinesep=0.3mm
		\begin{tabu}{|>{\RaggedRight\arraybackslash}p{2.4cm}|r|}
			\hline
			Variable & Value  \\\hline \hline
			$W_{N_1} = W_{N_2}$ & 548$\mu$m \\
			$W_{N_3}$ & 5$\mu$m \\
			$W_{N_4}$ & 290$\mu$m \\
			$W_{N_5} = W_{N_6}$ & 218$\mu$m \\
			$W_{N_7} = W_{N_8}$ &  30$\mu$m \\
			$W_{N_9} = W_{N_{10}}$ &  141$\mu$m \\
			$W_{N_{11}} = W_{N_{12}} = W_{N_{13}} = W_{N_{14}}$ &  84$\mu$m \\
			$W_{P_1} = W_{P_2}$ & 175$\mu$m\\
			$ W_{P_3} = W_{P_4}$ & 143$\mu$m \\
			$ W_{P_5} = W_{P_6}$ & 55$\mu$m \\
			$L_{N_1} = L_{N_2}$ & 8$\mu$m \\
			$L_{N_3} = L_{N_4} = L_{N_5} = L_{N_6} = L_{N_9} = L_{N_{10}} $ & 3$\mu$m \\
			$L_{N_7} = L_{N_8}$ & 2$\mu$m \\
			$L_{N_{11}} = L_{N_{12}} = L_{N_{13}} = L_{N_{14}}$ & 1$\mu$m \\
			$L_{P_1} = L_{P_2}$ & 2$\mu$m \\
			$L_{P_3} = L_{P_4} = L_{P_5} = L_{P_6}$ & 1$\mu$m\\
			\hline
		\end{tabu}
	}
	\qquad
	\subfloat[Symmetrical op-amp with high PSRR]{
		\label{tab:DimmensionsSymmetricalOpAmp}
		\tabulinesep=0.3mm
		\begin{tabu}{|>{\RaggedRight\arraybackslash}p{2.35cm}|r|}
			\hline
			Variable & Value  \\\hline \hline
			$W_{N_1} = W_{N_2}$ & 8$\mu$m \\
			$W_{N_3}$ & 56$\mu$m \\
			$W_{N_4} = W_{N_5}$ & 205$\mu$m \\
			$W_{N_6}$ &  460$\mu$m \\
			$W_{N_7}$ &  23$\mu$m \\
			$W_{P_1} = W_{P_2}$ & 5$\mu$m\\
			$ W_{P_3} = W_{P_4}$ & 15$\mu$m \\
			$ W_{P_5} = W_{P_6}$ & 35$\mu$m \\
			$ W_{P_7}$ & 287$\mu$m \\
			$L_{N_1} = L_{N_2}$ & 3$\mu$m \\
			$L_{N_3} = L_{N_7}$ & 6$\mu$m \\
			$L_{N_4} = L_{N_5}$ & 9$\mu$m \\
			$L_{N_6}$ & 1$\mu$m \\
			$L_{P_1} = L_{P_2} = L_{P_3} = L_{P_4} = L_{P_7}$ & 2$\mu$m \\
			$L_{P_5} = L_{P_6}$ & 2$\mu$m	\\
			$C_c$ &	4.5pF\\\hline
		\end{tabu}
	}
	\qquad
	\subfloat[Complementary op-amp]{
		\label{tab:RailToRailAmplifier}

		\tabulinesep=0.3mm
		\begin{tabu}{|>{\RaggedRight\arraybackslash}p{2.4cm}|r|}
			\hline
			Variable & Value  \\\hline \hline
			$W_{N_1}$ & 13$\mu$m \\
			$W_{N_2}$ & 306$\mu$m \\
			$W_{N_3} = W_{N_4}$ & 79$\mu$m \\
			$W_{N_5} = W_{N_6}$ & 39$\mu$m \\
			$W_{N_7} = W_{N_8}$ & 32$\mu$m \\
			$W_{N_9}$ & 13$\mu$m \\
			$W_{P_1}$ & 7$\mu$m \\
			$W_{P_2}$ & 168$\mu$m \\
			$W_{P_3} = W_{P_4}$ & 378$\mu$m \\
			$W_{P_5} = W_{P_6}$ & 66$\mu$m \\
			$W_{P_7} = W_{P_8}$ & 104$\mu$m \\
			$L_{N_1} = L_{N_2} = L_{N_9}$ & 4$\mu$m \\
			$L_{N_3} = L_{N_4}$ & 1$\mu$m \\
			$L_{N_5} = L_{N_6}$ & 5$\mu$m \\
			$L_{N_7} = L_{N_8}$ & 5$\mu$m \\
			$L_{P_1} = L_{P_2}$ & 3$\mu$m \\
			$L_{P_3} = L_{P_4}$ & 1$\mu$m \\
			$L_{P_5} = L_{P_6}$ & 3$\mu$m \\
			$L_{P_7} = L_{P_8}$ & 3$\mu$m \\\hline
		\end{tabu}
	}%
	\qquad
\end{table}

The instantiated equations and constraints are automatically given to the embedded constraint programming solver GeCode \cite{Gecode}. 
Several sizings are calculated for a topology
with a backtracking-search algorithm, which is based on branch-and-bound (BAB) methods.
The transistor dimensions were generated using a 0.25$\mu$m PDK. The supply voltage was 5V and the bias current 10$\mu$A.

Sizing values (Table \ref{tab:dimensionsOpAmp}) were generated for the circuits in Fig. \ref{fig:differentOpAmpToplogies}, using the specifications in Table \ref{tab:GenericSpecs}, \ref{tab:OtherSpecs}. The performance values calculated with the performance models and results from circuit simulation are included in these tables. The average deviations for all performance specifications are 9\% - 23\%. This meets the requirement of analog designers who expect a 20\% - 30\% deviation between the Shichman-Hodges model and full circuit simulation.
The largest deviation is obtained for the unity-gain bandwidth of the telescopic op-amp and of the symmetrical op-amp. It is overestimated and is one of few performance features that do not meet the specification. 
The unity-gain bandwidth depends linearly on the transconductance of the input transistor of the first stage \eqref{eq:UnitiGanBandwidthNormal}. This transconductance is often overestimated using the Shichman-Hodge model.

\begin{table}[tbp]
	\scriptsize\centering
	\caption{Performance values of the (a) telescopic op-amp (b) symmetrical op-amp with high PSRR }\label{tab:GenericSpecs}
	\setlength{\tabcolsep}{0.1cm}
	\begin{tabular}{|>{\raggedright\arraybackslash}m{2.5cm}||l|l||>{\centering\arraybackslash}m{0.6cm}|c||c|c||c|}
		\hline
		\multirow{2}{*}{Constraints} &\multicolumn{2}{c||}{ Spec.} & \multicolumn{2}{c||}{Sizing tool} & \multicolumn{2}{c||}{BSIM3v3} &\multirow{2}{0.9cm}{Average deviation}  \\\cline{2-7}
		& (a) & (b) & (a) & (b) &(a)&(b) & \\ \hline \hline
		Gate-area (10$^3$ $\mu$m$^2$) & $\leq$ 15 & $\leq$ 10 & 5.8 & 5.5 &- & -& - \\
		Quiescent power (mW) & $\leq$ 10 &  $\leq$ 15 & 5.8 & 4 & 6.1 & 4.5 &  13\% \\\hline		
		Max. common-mode input voltage (V) & $\geq$ 3  & $\geq$ 3& 3.3 & 4.3 & 4.4& 4.3 &-\\
		Min. common-mode input voltage (V) & $\leq$ 2 & $\leq$ 2 &0 & 0.8 & 0.1& 0.7 & -\\
		Max. output voltage (V) & $\geq$ 4 & $\geq$ 4 &  4.5 & 4.5 &4.5& 4.4 & - \\
		Min. output voltage (V) & $\leq$ 1 & $\leq$ 1&  0.3 & 0.1 & 0.2& 0.2 &- \\
		CMRR (dB) & $\geq$ 90 & $\geq$ 90& 130  & 95 & 146 & 142  &11\% / 33\% \\
		Unity-gain bandwidth (MHz) &$\geq$ 7 & $\geq$ 7 & 10 & 10.3 &\textcolor{red}{6.5} & \textcolor{red}{6.8} &53\% / 51\%\\\hline
		Open-loop gain (dB) & $\geq$ 80 & $\geq$ 80& 120 & 100 & 93& 97 &29\% / 3\% \\
		Slew rate ($\frac{\text{V}}{\mu \text{s}}$)& $\geq$ 15 & $\geq$ 10& 28 & 15 & 22 & 11&27\% / 36\% \\
		Phase Margin ($^\circ$) & $\geq$ 60 & $\geq$ 60& 60& 61 & 67 & \textcolor{red}{59}   & 10\% / 3\%  \\\hline
		Average deviation of all perf. values & -&-&- &-& - &-& 23\% / 23\%\\\hline
	\end{tabular}	
\end{table}

\begin{table}[tbp]
	\scriptsize\centering	\setlength{\tabcolsep}{0.1cm}
		\caption{Performance values of the (a) folded-cascode op-amp with CMFB (b) complementary op-amp  }\label{tab:OtherSpecs}
	\begin{tabular}{|>{\raggedright\arraybackslash}m{2.5cm}||l|l||>{\centering\arraybackslash}m{0.6cm}|c||c|c||c|}
		\hline
		\multirow{2}{*}{Constraints} &\multicolumn{2}{c||}{ Spec.} & \multicolumn{2}{c||}{Sizing tool} & \multicolumn{2}{c||}{BSIM3v3} &\multirow{2}{0.7cm}{Average deviation}  \\\cline{2-7}
		& (a) & (b)& (a) & (b) &(a)&(b) & \\ \hline \hline
		Gate-area (10$^3$ $\mu$m$^2$) & $\leq$ 15 & $\leq$ 5 & 13.4 & 4.5 &- & -& - \\
		Quiescent power (mW) & $\leq$ 15 &  $\leq$ 5& 10& 5 &{11} & 4.4 & 10\% / 14\% \\\hline
		Max. common-mode input voltage (V) & $\geq$ 3 & -& 4.5 & - & 4.4 & - &-\\
		Min. common-mode input voltage (V) & $\leq$ 2 &- &0.9&- & 1 &-  & -\\
		Max. output voltage (V) & $\geq$ 3.5& $\geq$ 3.5 & 4  & 3.5 & 4.1& 3.8  & - \\
		Min. output voltage (V) & $\leq$ 1&  $\leq$ 1.5 & 0.9 & 1.4 & 1 & 0.5 &- \\
		CMRR (dB) & $\geq$ 80& $\geq$  70  & 122& 133 & 118  & 136 &3\% / 2\%\\
		Unity-gain bandwidth (MHz) &$\geq$ 10 &$\geq$ 10&10 & 28 &10.5& 19&5\% / 47\%\\\hline
		Open-loop gain (dB) & $\geq$ 70 & $\geq$ 80 &75 & 84 & 71 & 86 &6\% / 2\%\\
		Slew rate ($\frac{\text{V}}{\mu \text{s}}$)& $\geq$ 15& $\geq$ 15 &24.5 & 23 & 19 & 20 &29\% / 15\%\\
		Phase Margin ($^\circ$) & $\geq$ 60  & $\geq$ 60 &82& 62 &83 & \textcolor{red}{57} &  1\% / 9\% \\\hline
		Average deviation of all perf. values. &- &-&-&- &- & - &9\% / 15\% \\\hline
	\end{tabular}
\end{table}

For the symmetrical and complementary op-amps, the phase margin requirement is not fulfilled. However, the deviation between the simulation and calculated value is very small, 3\% respectively 9\%. The equation-based model of the phase margin is quite accurate.

All other specifications are fulfilled by the calculated and simulated performance values.
\cite{ABNG20d} shows additional sizing results obtained with the HPEL. The paper presents a synthesis tool featuring thousands of different op-amp topology using the HPEL to evaluate op-amp topologies. Sizing results for 100 different topologies are compared. The average deviation is again between 20\% - 30\% and thus meets the expectation of designers.
Further simulation-based optimization may be performed on the circuit to improve the performance.

\begin{figure}[t]\centering
	\includegraphics[width=0.8\linewidth]{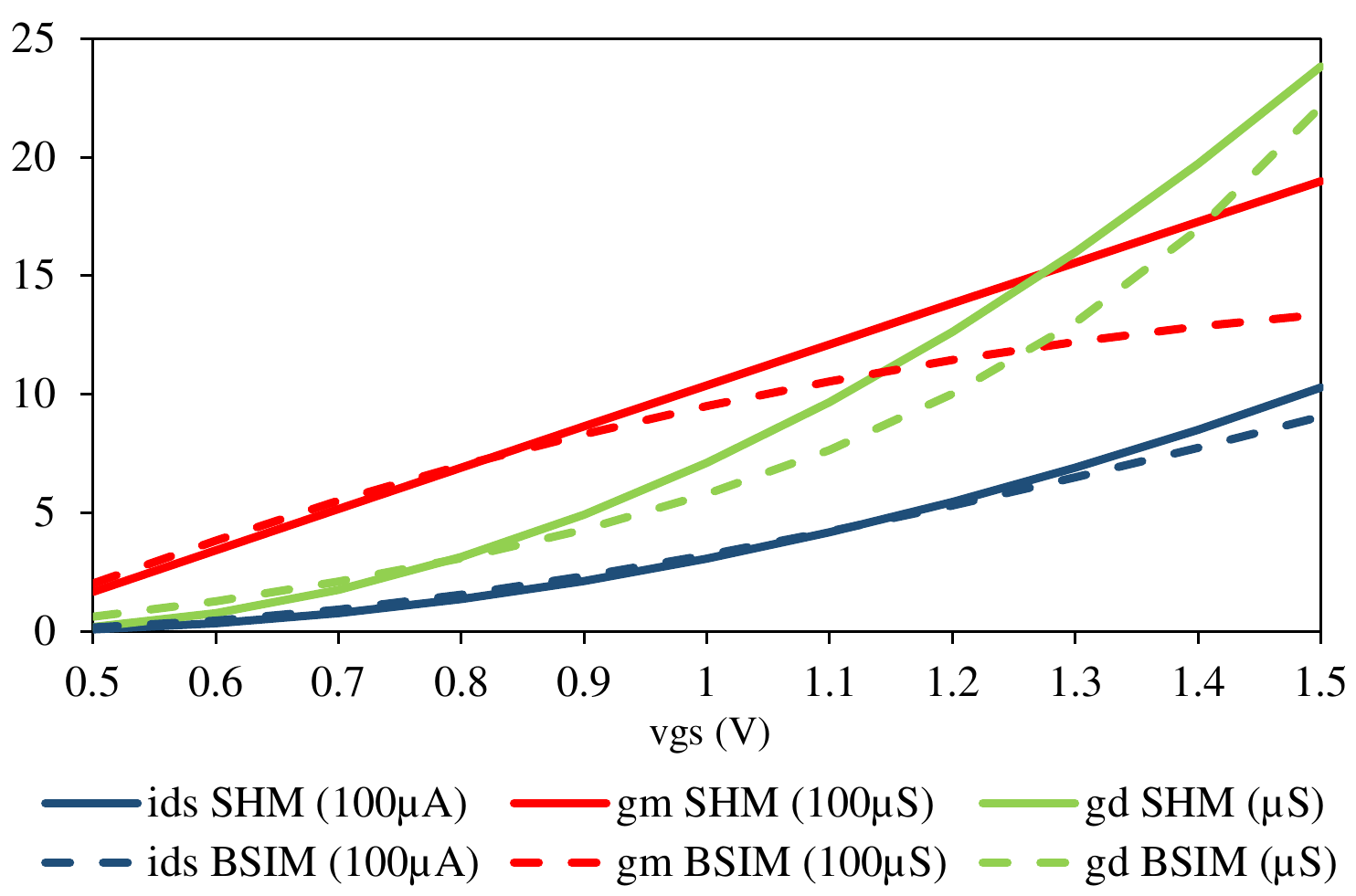}
	\caption{Transistor parameter for different $v_{GS}$-values}
	\label{fig:transistorParameterCurves}
\end{figure}

Fig. \ref{fig:transistorParameterCurves}  compares the  Shichman-Hodges model used in HPEL and the BSIM3v3 model used in  simulation. It shows the transconductance $gm$, the output conductance $gd$ and the drain-source current $i_{DS}$ of a transistor for different $v_{GS}$-values obtained with the two models. The transistor width and length are set to 10$\mu$m, 1$\mu$m respectively. The drain-source voltage was set to be $1.5$ V, such that the transistor operates in saturation with strong inversion, a common working region in analog circuits.
For small $v_{GS}$ values, the two transistor models correspond well. Higher $v_{GS}$-values lead to deviations. Keeping $v_{GS}$ small hence leads to accurate performance results using HPEL. Future work is on integrating more complex transistor models, such as the EKV model, into HPEL. The EKV model has a low complexity compared to BSIM3v3, but features a good accuracy in all transistor regions. 
Integrating more advanced transistor models makes the method also usable for modern technologies with small channel lengths. Other approaches integrate the gm/Id-method based on look-up tables in the sizing tool \cite{ANALOG2020}.

\section{Conclusions, Limitations, Outlook}\label{sec:Conclusion}

This paper presented a method to automate the set-up of an equation-based behavioral description of an op-amp and applied it to circuit sizing. A hierarchical performance equation library (HPEL) was developed, allowing the equations to be automatically set up based on a functional block analysis of the circuit. The created circuit model combines simulation and sizing as it uses KCL/KVL to simulate the currents and voltages in the circuit and performance equations to describe the circuit behavior suitable for sizing.
The analytical performance equation makes the usage of numerical performance evaluation during sizing unnecessary. 
The method is generic in the sense that new types of circuits are not considered by setting up the equations from scratch, but by extending the HPEL with the respective new functional blocks and equations.

For the method to be applicable, an analytical description of the circuit class has to be available. 
While for established circuit classes, e.g., \cite{MultiStage1,MultiStage2}, such descriptions exist, this may not be the case for a brandnew circuit class that just evolves.

Currently, the HPEL supports one- and two-stage op-amps with simple compensation structures. The method can be extended to advanced frequency compensation techniques as \cite{Capacitor1,Capacitor2} and multi-stage op-amps. As \cite{MultiStage1,MultiStage2} show, multi-stage op-amps are describable with analytical equations on a high level of abstraction, which can be added to the HPEL.
This, e.g., needs new equations to support the arising poles and zeros of nested compensation and feedback-loops, which can be developed based on the structural studies in \cite{MultiStage1,MultiStage2}.
Additionally, the concept of functional block description can be transferred to other analog circuit classes. This requires an extension  of the functional block decomposition method in \cite{ABNG20} as well as an extension of the HPEL and the corresponding algorithms. A cross-coupled pair for example is frequently part of an oscillator or comparator circuit. Its formalized structural description would be added to \cite{ABNG20}. Its behavioral equations would be added to the HPEL. 

\section*{Acknowledgment}
The authors would like to thank the Cusanuswerk for partly funding this work.

\ifCLASSOPTIONcaptionsoff
  \newpage
\fi

\bibliographystyle{IEEEtran}
\bibliography{custom_added_2}

\begin{IEEEbiography}[{\includegraphics[width=1in,height=1.25in,clip,keepaspectratio]{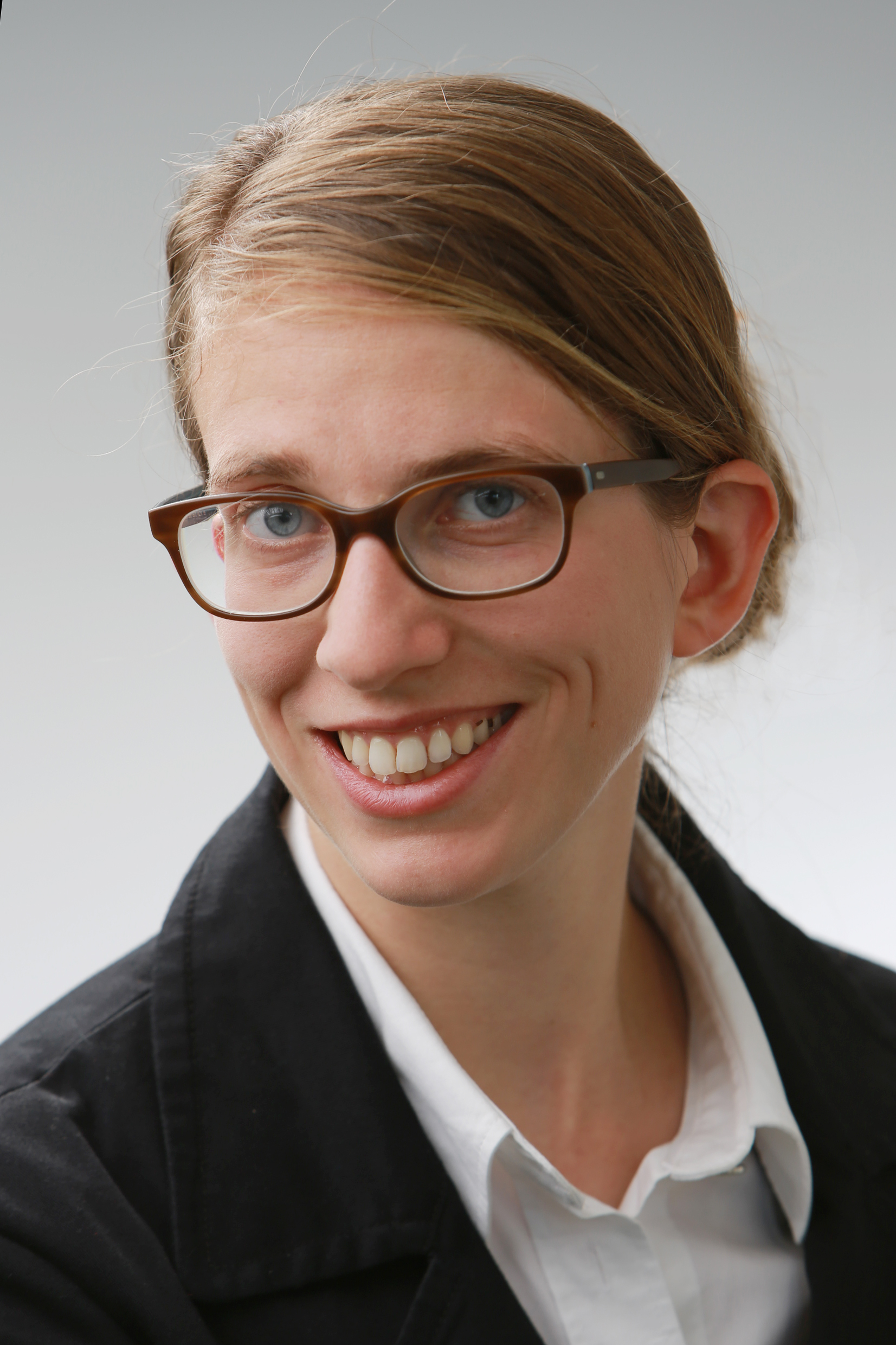}}]{Inga Abel}
	received her M. Sc. degree in Electrical and Computer Engineering from the Technical University of Munich, Germany 2018. She is currently working toward the Dr. Ing. degree at the Chair of Electronic Design Automation at the Technical University of Munich.
	Her research interests include the computer-aided design of analog integrated circuits and system, particularly the functional analysis, initial sizing and synthesis of analog circuits.
\end{IEEEbiography}
\begin{IEEEbiography}[{\includegraphics[width=1in,height=1.25in,clip,keepaspectratio]{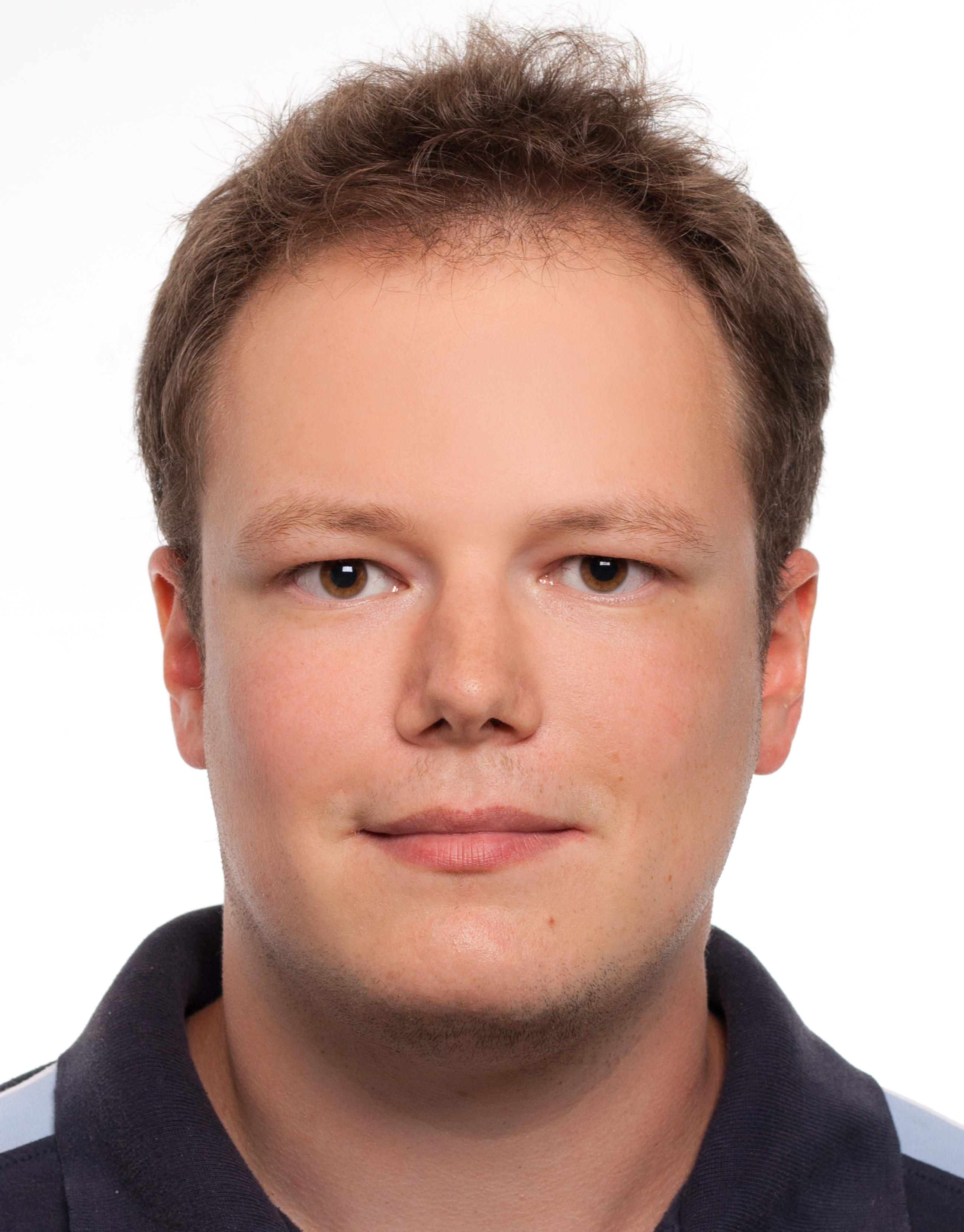}}]{Maximilian Neuner} received the B.Sc. and M.Sc. in Electrical and Computer Engineering from the Technical University of Munich (TUM) in 2013 and 2015 respectively. 
Since 2015, he is working towards the Ph.D. degree at the Chair of Electronic Design Automation at TUM.  
His research interests include the analysis, verification and synthesis of analog circuits. 
Mr. Neuner was a recipient of the 2015 Kurt Fischer Master Award for his master thesis from the Eikon e.V.
\end{IEEEbiography}
\begin{IEEEbiography}[{\includegraphics[width=1in,height=1.25in,clip,keepaspectratio]{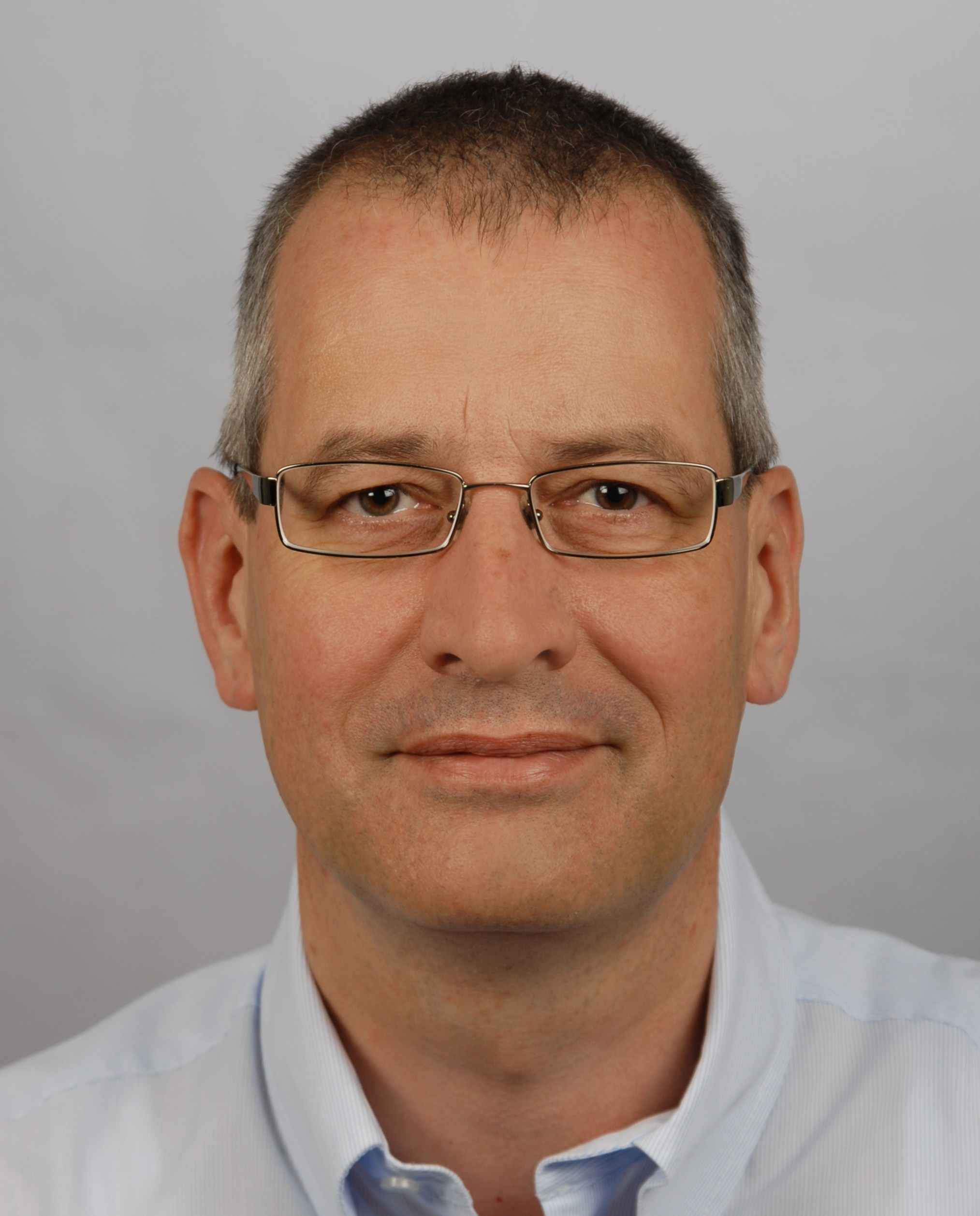}}]{Helmut E. Graeb}
	(M’02-SM’03-F`14) received the Dipl.-Ing., Dr.-Ing., and Habilitation degrees in electrical engineering from the Technical University of Munich (TUM), Munich, Germany, in 1986, 1993 and 2008, respectively. He was with Siemens Corporation, Munich, from 1986 to 1987, where he was involved in the design of DRAMs. Since 1987, he has been with the Chair of Electronic Design Automation, TUM, where he has been the Head of a research group since 1993. He has published more than 200 papers, six of which were nominated for best papers at the Design Automation Conference (DAC), the International Conference on Computer-Aided Design (ICCAD), and the Design, Automation and Test in Europe (DATE) conference. 
His research area is the design automation for analog and mixed-signal circuits. 
\end{IEEEbiography}

\end{document}